\documentclass{aa}
\usepackage{amsmath}
\usepackage{graphicx}
\usepackage{natbib}

\bibpunct{(}{)}{;}{a}{}{,}
\begin{document}

   \title{Thermodynamic evolution of cosmological baryonic gas:\\
   I. Influence of non-equipartition processes}

   \author{St\'ephanie Courty
          \inst{1,2}
          \and
          Jean-Michel Alimi\inst{1}}

   \offprints{S. Courty}
   
   \institute{Laboratoire de l'Univers et de ses Th\'eories, CNRS UMR
     8102, Observatoire de Paris-Meudon, 5 place Jules Janssen, 92195
     Meudon, France\\ \email{jean-michel.alimi@obspm.fr} \and Present
     address: Science Institute, University of Iceland, Dunhagi 3, 107
     Reykjavik, Iceland \\ \email{courty@raunvis.hi.is} }

   \date{Received date / Accepted date}

   \abstract{Using N-body/hydrodynamic simulations, the influence of
     non-equipartition processes on the thermal and dynamical
     properties of cosmological baryonic gas is investigated. We focus
     on a possible departure from equilibrium between electrons, ions
     and neutral atoms in low temperature ($10^4$--$10^6$ K) and
     weakly ionized regions of the intergalactic medium. The
     simulations compute the energy exchanges between ions, neutrals
     and electrons, without assuming thermal equilibrium. They include
     gravitation, shock heating and cooling processes, and follow
     self-consistently the chemical evolution of a primordial
     composition hydrogen-helium plasma without assuming collisional
     ionization equilibrium. At high redshift, a significant fraction
     of the intergalactic medium is found to be warmer and weakly
     ionized in simulations with non-equipartition processes than in
     simulations in which the cosmological plasma is considered to be
     in thermodynamic equilibrium.  With a semi-analytical study of
     the out of equilibrium regions we show that, during the formation
     of cosmic structures, departure from equilibrium in accreted
     plasma results from the competition between the atomic cooling
     processes and the elastic processes between heavy particles and
     electrons. Our numerical results are in agreement with this
     semi-analytical model. Therefore, since baryonic matter with
     temperatures around $10^4$ K is a reservoir for galaxy formation,
     non-equipartition processes are expected to modify the properties
     of the objects formed.

     \keywords{cosmology: theory -- large-scale structure of the
       universe -- intergalactic medium -- galaxies: formation -- 
       hydrodynamics} }

     \titlerunning{Thermodynamic evolution of cosmological baryonic gas}
     \authorrunning{St\'ephanie Courty \& Jean-Michel Alimi}

     \maketitle
%

\section{Introduction}

In the basic picture of galaxy formation, baryons fall into dark
matter potential wells located along the network of sheets and
filaments formed by gravitational instability
\citep{Sunyaev1972,Rees77, Silk77, White78}. In this hierarchical
picture, galaxy clusters result from the large scale collapse of gas
and from mergers of subunits. These massive structures ($M>10^{14}$
M$_{\sun}$) are mainly present at low redshift ($z<2$). Mergers and
accretion of matter involve hydrodynamical shocks raising the
intracluster medium temperature to more than $10^7$ K. On the other
hand in galaxy-size structures, shocks and gravitational compression
do not heat the intergalactic medium to such high temperatures and the
gas can cool to a few times $10^4$ K and concentrate to form
galaxies. The intergalactic medium (IGM) refers to the cosmological
gas in gravitationally bound structures\footnote{In large scale
structure formation simulations, the word structure refers to any
potential well created by the dark matter. This sense is then more
general than referring only to galaxies, since in such potential wells
a number of galaxies can form. Hence the cosmological gas in
gravitationally bound structures, the medium inside potential wells,
is often referred to as intergalactic medium in the literature
\citep{Cen1993, Dave2001}.}. The hierarchical scenario causes the IGM
to separate into two phases: the hot intra-cluster medium at $T>10^6$
K and the cold intergalactic medium at a few times $10^4$ K
\citep{Cen93, Evrard}. But the process of accretion and cooling of
baryons inside bounded structures is still not well understood and a
study of the thermodynamic properties of the IGM is crucial to our
understanding of galaxy formation.

The cosmological plasma accreted in bound structures acquires thermal
energy through gravitational compression and shock heating. The
increase in temperature heats preferentially heavy particles and owing
to the large difference in mass between heavy particles and electrons
the energy transfer between both species is poorly efficient
\citep{ZR}. The outer regions of bound structures then consist of a
non-equilibrium two temperature gas. In this paper we introduce
non-equipartition processes between ions, neutrals and electrons of
the cosmological plasma in numerical simulations of large scale
structure formation in order to investigate their influence on the
thermodynamic properties of baryons. The results are compared to
simulations in which equipartition processes are forced. The
simulations are performed with an Eulerian N-Body/hydrodynamical code
including gravitation, shock heating and radiative cooling processes.

Depending on the depth of the potential wells, the temperature of the
accreted baryonic matter ranges from $10^4$ to $10^8$ K. At high
temperatures, the accreted matter is totally ionized behind shock
fronts and departure from equilibrium between ions and electrons is
due to the quite long energy exchange timescale between these species.
The thermal decoupling between ions and electrons in galaxy clusters
has been considered by several authors \citep{Fox, Chieze98,
Takizawa_a, Takizawa_b}. Indeed \cite{Chieze98} show that temperatures
of these two species can be significantly different, by a factor of
$\sim 3$, in the outer regions of galaxy clusters. Our large scale
simulations do show such temperature differences in massive structures
at low redshift.\\

The simulations also show departure from equilibrium in structures
much less massive than galaxy clusters.  These structures are
characterized by a warm plasma with heavy particle temperature between
$10^4$ and $10^6$ K and electron temperature at a few times $10^4$ K.
That plasma is also weakly ionized.  We show that the fraction of warm
IGM dramatically increases at high redshift because of the larger
number of low mass structures, formed in hierarchical models.
Departure from equilibrium is not only due to Coulomb interactions
between charged particles, but also due to mechanisms of short-range
forces between electrons and neutral particles \citep{Petschek,
Shchekinov}.  In such low density IGM, equilibration timescales
between these species are short compared to the Hubble time and we
propose a semi-analytical model to explain them. This model suggests
some insights into the thermodynamical story of cosmological baryonic
gas.  \\

In this paper we focus on the low mass structure case and examine in
detail the thermodynamic properties of the baryons. We show how
non-equipartition processes, inside bound structures, at high redshift
and before the reionization epoch, yields a large fraction of the IGM
to be warm (between $10^4$--$10^6$ K) and we discuss the implications
for galaxy formation. The paper is organized as follows.  Section
\ref{simul} presents the numerical code and simulations. In section
\ref{prop} influence of non-equipartition processes on the
thermodynamic properties of baryons are analyzed. Section
\ref{discuss} gives, with the aid of a semi-analytical model, the
physical origin of departure from equilibrium in the low mass
structure case. Finally Section \ref{csq} discusses the cosmological
implications for galaxy formation.

\section{Simulations}
\label{simul}

Simulations are performed with a 3 dimensional N-body/hydrodynamical
code coupling a Particle-Mesh code to compute gravitational forces
with an Eulerian (pseudo-Lagrangian) hydrodynamical code
\citep{Teyssier98}. The gravitational potential is solved from the
dark matter and the gas density fields. The main feature of the code
is the computation of non-equipartition processes between ions,
neutrals and electrons. Each species of the cosmological plasma has
its own specific internal energy, $e_\mathrm i$, $e_\mathrm n$,
$e_\mathrm e$, satisfying the energy conservation equation:
\begin{equation}
\Big( \rho \frac{\mathrm{D}e}{\mathrm{D}t} = -p(\boldsymbol{\nabla}
\cdot \boldsymbol{v}) + \Psi - \Lambda_\mathrm{net} \Big)_x,
\end{equation}
where $\rho_x$ is the density, $e_x$ the specific internal energy,
$p_x=n_xkT_x$ the partial pressure and $n_x$ the number density for
the $x$ species. We do not distinguish the velocity $v$ for the three
species. $\Psi$ is a dissipative term due to shock heating and
$\Lambda_\mathrm{net}$ is a net cooling term containing the radiative
cooling rates (heating and cooling processes) and the energy exchange
rates between species. We now briefly discuss the two terms of the
energy equation, shock heating and elastic or inelastic processes,
computed separately in the code. \\

Shock heating is treated using the artificial viscosity method
\citep{VNR}. To reduce numerical dissipation, the contribution of
shocks to the energy is computed from a pseudo-entropy equation:
\begin{equation}
\label{DsDt}
\Big( \frac{\mathrm{D}s}{\mathrm{D}t} = \frac{s}{e} V_\mathrm s \Psi \Big)_x,
\end{equation}
where the pseudo-entropy $s_x$ for the species $x$ is defined by
$e_xV_\mathrm s^{\gamma-1}$ with gamma being the ratio of specific
heats and $V_\mathrm s$ the specific volume. The viscous dissipation
function is:
\begin{equation}
\label{Qvterm}
\Psi = -Q_\mathrm v (\vec{\nabla} \cdot \vec{v}),
\end{equation}
where the term $Q_\mathrm v$ equals
$-p(C_1\varepsilon+C_2\varepsilon^2)$ with $\varepsilon = \Delta x
\vec{\nabla} \cdot \vec{v}/c_\mathrm s$. $\Delta x$ is the physical
length of the cell and $c_\mathrm s$ the sound velocity. The constants
$C_1$ and $C_2$ are determined experimentally and set to 1 in the
simulations (the sensitivity to these constants are discussed in
section \ref{Qv}). The advantage of solving an equation for the
pseudo-entropy rather than the internal energy is that an adiabatic
flow remains strictly adiabatic.\\

The net cooling term is expressed as the contribution of three terms:
\begin{equation}
\Lambda_\mathrm{net} = \mathcal{H} + \Lambda + Q_\mathrm{exch},
\end{equation}
where $\mathcal{H}$ is a heating term for example due to
photoionization processes (see section \ref{csq}), $\Lambda$ is the
sum of the cooling rates and $Q_\mathrm{exch}$ denotes the energy
exchange between interacting species.

Radiative cooling processes considered here are: collisional
excitation, collisional ionization, recombination, bremsstrahlung and
Compton scattering. Ionization and heat input from the ultraviolet
radiation background are not included (this point is discussed in
section \ref{csq}). The computation of cooling rates needs to follow
the chemical evolution of a primordial composition hydrogen-helium
plasma with $n_\mathrm H$ and $n_\mathrm{He}$ the number densities of
hydrogen and helium, respectively. It must be noted that collisional
ionization equilibrium is not assumed here. The density evolution for
the six species of the plasma ($\mathrm H^0$, $\mathrm{He}^0$,
$\mathrm H^+$, $\mathrm{He}^+$, $\mathrm{He}^{++}$, $\mathrm e$) are
solved with:
\begin{eqnarray}
\label{pcH0}
-\beta_{\mathrm H^0} n_\mathrm e n_\mathrm {H^0} + \alpha_{\mathrm
 H^+} n_\mathrm e n_{\mathrm H^+} &=& \frac{\partial n_{\mathrm
 H^0}}{\partial t},\\
\label{pcHe0}
-\beta_{\mathrm{He}^0} n_\mathrm e n_{\mathrm{He}^0} +
 \alpha_{\mathrm{He}^+} n_\mathrm e n_{\mathrm{He}^+} &=&
 \frac{\partial n_{\mathrm{He}^0}}{\partial t},\\
\label{pcHe++}
\beta_{\mathrm{He}^+} n_\mathrm e n_{\mathrm{He}^+} -
\alpha_{\mathrm{He}^{++}} n_\mathrm e n_{\mathrm{He}^{++}} &=&
\frac{\partial n_{\mathrm{He}^{++}}}{\partial t},
\end{eqnarray}
where $\beta_i$ and $\alpha_i$ are the ionization and recombination
rates presented in Table~\ref{alphabetarates}, and the conservation
equations:
\begin{eqnarray}
n_{\mathrm H^0} + n_{\mathrm H^+} &=& n_\mathrm H, \\
\label{conserHe}
n_{\mathrm{He}^0} + n_{\mathrm{He}^+} + n_{\mathrm{He}^{++}} &=&
n_\mathrm {He}, \\ n_{\mathrm H^+} + n_{\mathrm{He}^+} +
2n_{\mathrm{He}^{++}} &=& n_\mathrm e.
\end{eqnarray}
The number densities of the six species allow computation of cooling
rates and since the numerical code allows ions, neutrals and electrons
to have their own temperature, cooling rates depend on electron
temperature and are detailed in Table~\ref{coolrates}.\\

Energy transfers between the three populations are expressed by:
\begin{equation}
Q_{x_1,x_2} = -Q_{x_2,x_1} = \frac{(n_{x_1}/n_{x_2})e_{x_2} -
e_{x_1}}{\tau_{x_1,x_2}},
\end{equation}
where $\tau_{x_1,x_2}$ is the characteristic timescale of interaction
between $x_1$ and $x_2$ species. The energy exchange between electrons
and ions is due to Coulomb interactions and the exchange timescale is
\citep{Spitzer}:
\begin{equation}
\label{teion}
\tau_\mathrm{e,ion} (\mathrm{s}) \approx 251.5 \ \frac{A \ T_\mathrm
e^{3/2}}{n_\mathrm e \ \mathrm{ln}(\Lambda)},
\end{equation}
where $A$ the molecular weight, $T_\mathrm e$ the electronic
temperature and $\mathrm{ln}(\Lambda)$ the Coulomb logarithm. The
energy exchange between electrons and neutrals is due to short-range
forces and, using the classical ``hard reflecting sphere''
cross-section $\sigma_\mathrm{e,neut} \approx 10^{-15} \
\textrm{cm}^2$ \citep{Draine}, the exchange timescale can be written
as \citep{ZR}:
\begin{equation}
\label{teneut}
\tau_\mathrm{e,neut} (\mathrm{s}) \approx \frac{A}{4.5 \times 10^{-13}
\ n_\mathrm n \ T_\mathrm e^{1/2}}.
\end{equation}
The energy exchange timescale between ions and neutrals is very short
compared to the other timescales above. It is worth noting that the
term $Q_\mathrm{exch}$ represents an energy gain for electrons and an
energy loss for the heavy particles.\\

\begin{table}
\begin{center}
\caption{\label{alphabetarates} Ionization and recombination rates
expressed in $\mathrm{s}^{-1} \ \mathrm{cm}^3$ (from \cite{Black} and
\cite{Cen92}) with $T_{\mathrm en}=T_\mathrm e/10^n$ K the electron
temperature.}
\begin{tabular}{l}
Ionization rates:\\
\label{betaH0}
$\beta_{\mathrm H^0} = 5.85\times10^{-11} (1+T_{\mathrm
e5}^{1/2})^{-1} \mathrm e^{(-157809.1/T_\mathrm e)} T_\mathrm e^{1/2}$\\
$\beta_{\mathrm{He}^0} = 2.38\times10^{-11} (1+T_{\mathrm
e5}^{1/2})^{-1} \mathrm e^{(-285335.4/T_\mathrm e)} T_\mathrm e^{1/2}$\\
$\beta_{\mathrm{He}^+} = 5.68\times10^{-12} (1+T_{\mathrm
e5}^{1/2})^{-1} \mathrm e^{(-631515/T_e)} T_\mathrm e^{1/2}$\\ \\
Recombination rates:\\ $\alpha_{\mathrm H^+} = 8.4\times10^{-11}
(1+T_{\mathrm e6}^{0.7})^{-1} T_\mathrm e^{-1/2} T_{\mathrm
e3}^{-0.2}$\\ $\alpha_{\mathrm{He}^+} = 1.5\times10^{-10} T_\mathrm
e^{0.6353}$\\
\label{alphaHepp}
$\alpha_{\mathrm{He}^{++}} = 3.36\times10^{-10} (1+T_{\mathrm
e6}^{0.7})^{-1} T_\mathrm e^{-1/2} T_{\mathrm e3}^{-0.2}$\end{tabular}
\end{center}
\end{table}

\begin{table}
\begin{center}
\caption{\label{coolrates} Cooling rates expressed in $\mathrm{erg} \
\mathrm{s}^{-1} \ \mathrm{cm}^{-3}$ (from \cite{Black} with
modifications introduced by \cite{Cen92} for temperatures exceeding
$10^5$ K) with $T_\mathrm{en}=T_\mathrm e/10^n$ K the electron
temperature, $g_\mathrm{ff}$ the Gaunt factor taken to 1.5, and
$T_\mathrm{bg}$ the cosmic microwave background temperature,
$T_\mathrm{bg}=2.7(1+z)$.}
\begin{tabular}{l}
Collisional excitation: \\
$\Lambda_{\mathrm H^0} = 7.5\times10^{-19} (1+T_{\mathrm e5}^{1/2})^{-1} \mathrm e^{(-118348/T_\mathrm e)} n_\mathrm e n_{\mathrm H^0}$\\
$\Lambda_{\mathrm{He}^+} = 5.54\times10^{-17} (1+T_{\mathrm e5}^{1/2})^{-1} \mathrm e^{(-473638/T_\mathrm e)} T_\mathrm e^{-0.397} n_\mathrm e n_{\mathrm{He}^+}$\\
\\
Collisional ionization:\\
$\Lambda_{\mathrm H^0} = 1.27\times10^{-21} (1+T_{\mathrm e5}^{1/2})^{-1} \mathrm e^{(-157809.1/T_\mathrm e)} T_\mathrm e^{1/2} n_\mathrm e n_{\mathrm H^0}$\\
$\Lambda_{\mathrm{He}^0} = 9.38\times10^{-22} (1+T_{\mathrm e5}^{1/2})^{-1} \mathrm e^{(-285335.4/T_\mathrm e)} T_\mathrm e^{1/2} n_\mathrm e n_{\mathrm{He}^0}$\\
$\Lambda_{\mathrm{He}^+} = 4.95\times10^{-22} (1+T_{\mathrm e5}^{1/2})^{-1} \mathrm e^{(-631515/T_\mathrm e)} T_\mathrm e^{-0.397} n_\mathrm e n_{\mathrm{He}^+}$\\
\\
Recombination:\\
$\Lambda_{\mathrm H^+} = (3/2)kT_\mathrm e\times 8.4\times10^{-11} (1+T_{\mathrm e6}^{0.7})^{-1} T_\mathrm e^{1/2} T_{\mathrm e3}^{-0.2} n_\mathrm e n_{\mathrm H^+}$\\
$\Lambda_{\mathrm{He}^+} = (3/2)kT_\mathrm e\times 1.5\times10^{-10} T_\mathrm e^{0.3647} n_\mathrm e n_{\mathrm{He}^+}$\\
$\Lambda_{\mathrm{He}^{++}} = (3/2)kT_\mathrm e\times 3.36\times10^{-10} (1+T_{\mathrm e6}^{0.7})^{-1} T_\mathrm e^{1/2} T_{\mathrm e3}^{-0.2} n_\mathrm e n_{\mathrm{He}^{++}}$\\
\\
Bremsstrahlung:\\
$\Lambda_\mathrm{ff} = 1.42\times10^{-27} g_\mathrm{ff} T_\mathrm e^{1/2} (n_{\mathrm H^+}+n_{\mathrm{He}^+}+4n_{\mathrm{He}^{++}})n_\mathrm e$ \\
\\
Compton interaction:\\
$\Lambda_\mathrm C = 5.65\times10^{-36}(1+z)^4(T_\mathrm e-T_\mathrm{bg})n_\mathrm e$\\
\end{tabular}
\end{center}
\end{table}

\begin{figure*}
\begin{center}
\includegraphics[height=6cm]{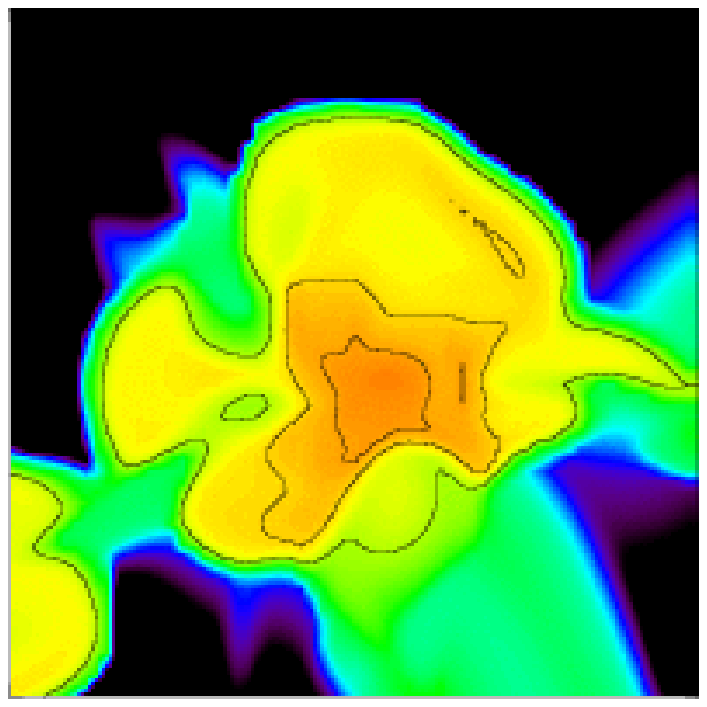}
\includegraphics[height=6cm]{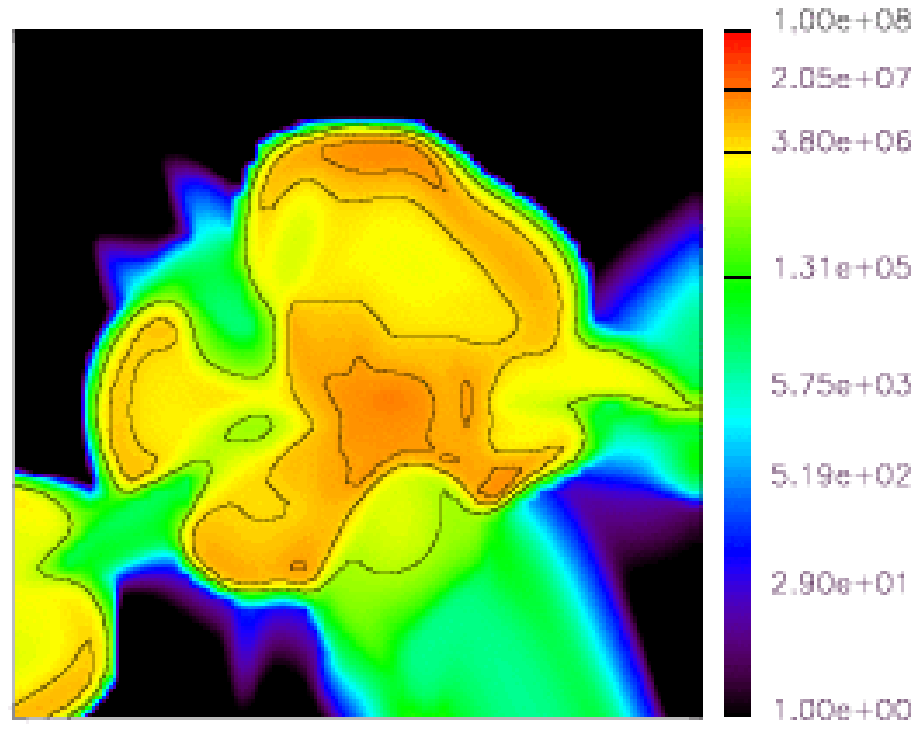}
\caption{Isocontours of electron (left panel) and heavy particle
(right panel) temperatures for a high mass structure extracted from
the large scale $S_{3\mathrm T}$ simulation including
non-equipartition processes (the slices are 8 comoving $h^{-1}
\textrm{Mpc}$ on a side and contour levels are: $10^6, 5.10^6, 10^7,
2.10^7, 5.10^7, 10^8$ K).}
\label{amasS3T}
\end{center}
\end{figure*}

The simulations are performed for a flat $\Lambda$-dominated cold dark
matter model defined by $\Omega_\mathrm m=0.3$,
$\Omega_{\Lambda}=0.7$, $\Omega_\mathrm bh^{2}=0.02$,
$h=(H_0/100)=0.7$. The transfer function is taken from \cite{Bardeen}
with a shape parameter from \cite{Sugiyama}. We use the normalization
on COBE data \citep{Bunn} giving at $R=8 \ h^{-1} \textrm{Mpc}$ a
filtered dispersion $\sigma_8=0.91$. The computational cubic volume is
of 16 comoving $h^{-1} \textrm{Mpc}$ on a side with $N_\mathrm
g=256^3$ grid cells and $N_\mathrm p=256^3$ dark matter particles
allowing a spatial resolution of 62.5 $h^{-1} \textrm{kpc}$.  The dark
matter particle mass is $M_\mathrm{dm}=2.51\times 10^7$ M$_{\sun}$ and
the gas mass resolution is $M_\mathrm{bm}=3.87\times 10^6$ M$_{\sun}$.
In Section \ref{high_strt}, the simulations are 32 comoving $h^{-1}
\textrm{Mpc}$ on a side with other parameters similar.

We refer to simulations with non-equipartition processes as
$S_{3\mathrm T}$ and the results are compared with simulations in
which equipartition processes are forced. In that case the three
species have the same temperature and these simulations are denoted
$S_{1\mathrm T}$.

\section{Thermodynamic properties of baryons}
\label{prop}

Accretion of baryons into potential wells created by dark matter
involves gravitational compression and hydrodynamical shocks.
Dissipative processes convert the kinetic energy of the cosmological
plasma into thermal energy and the increase in temperature depends on
the particle mass (eq. \ref{Ta0}). Owing to the large difference of
mass between electrons and heavy particles (ions and neutral atoms),
the energy transfer between these species is poorly efficient, leading
to an out of equilibrium plasma. The next two subsections describe the
thermodynamic evolution of the cosmological plasma, first inside high
mass structures comparable to galaxy clusters and, secondly inside low
mass structures comparable to proto-galaxies.

\subsection{High mass structures or totally ionized IGM}
\label{high_strt}

\begin{figure*}
\begin{center}
\includegraphics[height=6cm]{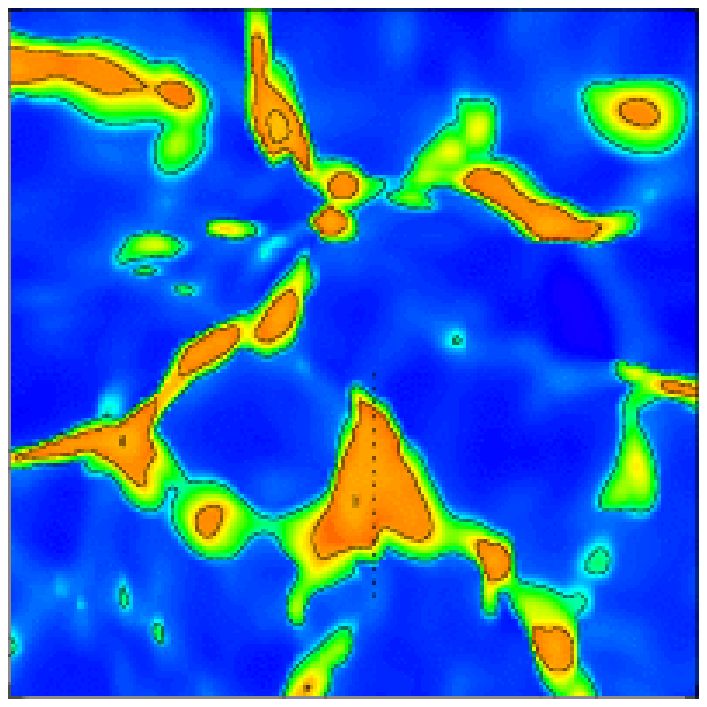}
\includegraphics[height=6cm]{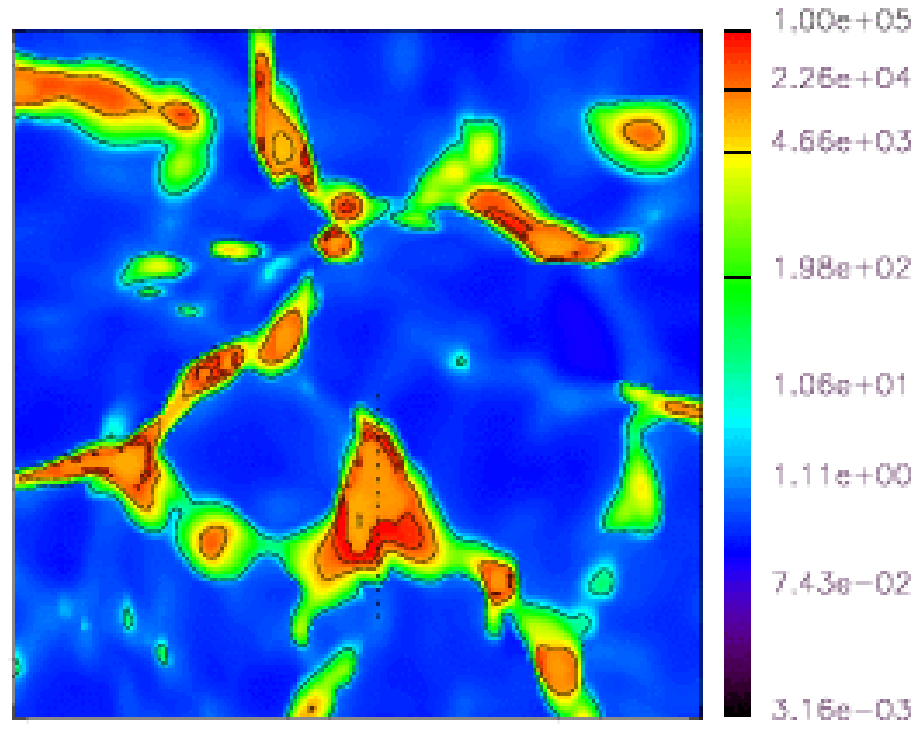}
\caption{Isocontours of the gas temperature in the $S_{1\mathrm T}$
  simulation (left panel) and the heavy particle temperature in the
  $S_{3\mathrm T}$ simulation (right panel) at $z=4$ (the slices are 8
  comoving $h^{-1} \textrm{Mpc}$ on a side and contour levels are:
  $10., 10^4, 4.10^4$ K).}
\label{carteTS3T}
\end{center}
\end{figure*}

Figure~\ref{amasS3T} focuses on a high mass structure in the
$S_{3\mathrm T}$ simulation and displays electron and ion temperature
distributions.  This cluster-like structure is extracted from a large
scale field at $z=0$.  Isocontours illustrates shock heated baryonic
matter at temperatures around $10^7$--$10^8$ K which is typically the
hot intra-cluster medium temperature. Indeed, galaxy clusters result
from the merger of sub-units of matter inside deep potential wells
created by the large scale collapse of dark matter. Hydrodynamic
shocks induced by mergers heat the heavy particles to high
temperatures and in the outer regions, just behind the shock fronts,
the electron temperature is significantly lower than the ion
temperature.

In the case of discontinuities, the post-shock temperature for each
species of the plasma is computed with the conservation equations
which connect impulsion, $\rho u$, pressure, $p$, and specific
enthalpy, $h=H/\rho$, on both sides of the thin compression
region. Ions and neutral particles are grouped under the term heavy
particles with the subscript $\mathrm{HP}$. Pre-shock quantities in
the unperturbed region are denoted with the subscript $0$ and no
subscript is reserved for post-shock quantities. The Rankine-Hugoniot
relations, written with the compression ratio $\omega = \rho_0/\rho$,
are:
\begin{eqnarray} 
\rho u &=& \rho_0 u_0, \\
\label{conservp}
p &=& \rho_0 u_0^2 (1-\omega) + p_0,\\
\label{conservh}
h &=& \frac{1}{2} u_0^2 (1-\omega^2) + h_0.
\end{eqnarray} 
The specific enthalpy for the plasma is expressed by:
\begin{equation}
h = \frac{1}{\rho} \left( \frac{5}{2}n_\mathrm{HP}kT_\mathrm{HP} + \frac{5}{2}n_\mathrm ekT_\mathrm e \right),
\end{equation}
with $n_\mathrm e$ and $n_\mathrm{HP}$ the number densities of the
electrons and the heavy particles, respectively, and $T_\mathrm e$ and
$T_\mathrm{HP}$ the electron and the heavy particle temperatures,
respectively. The increase in temperature is then calculated by
assuming that the number density per mass unit for each species
($n_x/\rho$) is constant across the discontinuity:
\begin{equation}
\label{DeltaTL}
T_\mathrm{HP} - T_{\mathrm{HP}_0} = \frac{n_{\mathrm
e_0}}{n_{\mathrm{HP}_0}} (T_{\mathrm e_0}-T_\mathrm e) + \frac{\rho_0
u_0^2 (1-\omega^2)}{5kn_{\mathrm{HP}_0}},
\end{equation}
The electron gas undergoes adiabatic compression across the
discontinuity and the electron temperature increases by a factor of
$1/\omega^{\gamma-1}$ \citep{ZR}. The variation $(T_{\mathrm
e_0}-T_\mathrm e)$ can then be neglected in equation
(\ref{DeltaTL}). Moreover for strong shocks, $1/\omega \approx 4$, and
the last term becomes $\approx (3/16)(\rho_0
\mathcal{D}^2/(kn_{\mathrm{HP}_0}))$. By introducing the molecular
weight of the gas $A$, the increase in temperature for heavy particles
becomes:
\begin{eqnarray}
\label{Ta0}
\Delta T_\mathrm{HP} &\approx& \frac{3}{16} \frac{A m_\mathrm p}{k} u_0^2,\\
\label{Ta1}
&\approx& 2.2 \times 10^5 \ \textrm{K} \left( A \cdot
\frac{\mathcal{D}}{100 \ \textrm{km \ s}^{-1}}
\right)^2. \end{eqnarray}

As gravitational compression and shocks heat preferentially the heavy
particles, the temperature of the two species can differ significantly
in an extended relaxation layer during a timescale which depends on
the energy transfer between particles. Plasma in galaxy clusters is
fully ionized and for typical values of density and temperature in
outer regions, an estimate of the characteristic timescale for the
energy exchange between electrons and ions gives (see eq.
\ref{teion}):
\begin{equation}
\label{teion2}
\tau_\mathrm{e,ion} \approx 9.4 \times 10^9 \ \mathrm{yr} \ \frac{
(T_\mathrm e/5.10^7 \ \mathrm{K})^{3/2}}{(n_\mathrm e/10^{-5} \
\mathrm{cm}^{-3}) (\mathrm{ln} (\Lambda)/30)}.
\end{equation}

This relaxation timescale is of the order of the Hubble time
$\tau_\mathrm H$, and explains the departure from equilibrium in the
outer regions such as seen in Fig.~\ref{amasS3T}. For higher density,
this timescale decreases and the equilibrium can be rapidly recovered
between both species: the distributions in Fig.~\ref{amasS3T} show
that, in the structure center, both temperatures are similar. A number
of earlier studies show such differences between temperatures in
galaxy cluster high-resolution simulations based on Hoffman-Ribak
initial conditions \citep{Chieze98, Takizawa_a, Takizawa_b}. Here we
reach similar conclusions but obtained using simulations of large
scale structure formation.

Departure from equilibrium in the outer regions of a massive galaxy
cluster is a first illustration of the influence of non-equipartition
processes on the IGM. However, in hierarchical models massive galaxy
clusters exist mainly at low redshifts and an out of equilibrium
plasma at temperature higher than $10^7$ K is only expected at these
redshifts.  We now analyze the influence of non-equipartition
processes on the cosmological plasma over the evolution of the
universe, at epochs when a larger fraction of baryonic matter is
contained in low mass structures.

\subsection{Low mass structures or weakly ionized IGM}
\label{low_strt}

\begin{figure}
\begin{center}
\includegraphics[height=5.7cm]{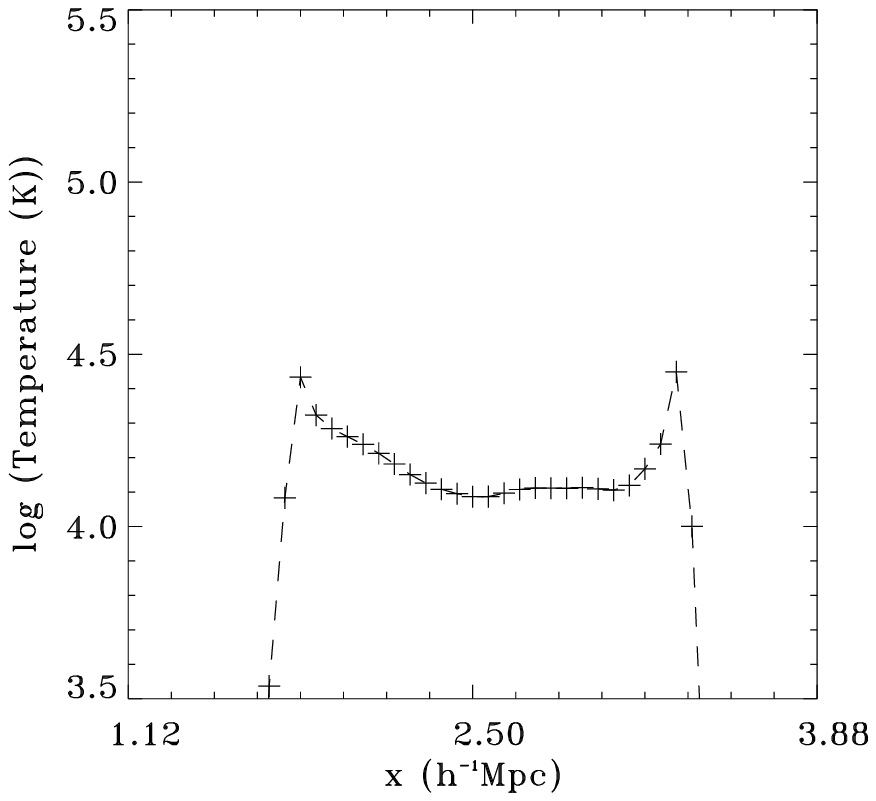}\\
\includegraphics[height=5.7cm]{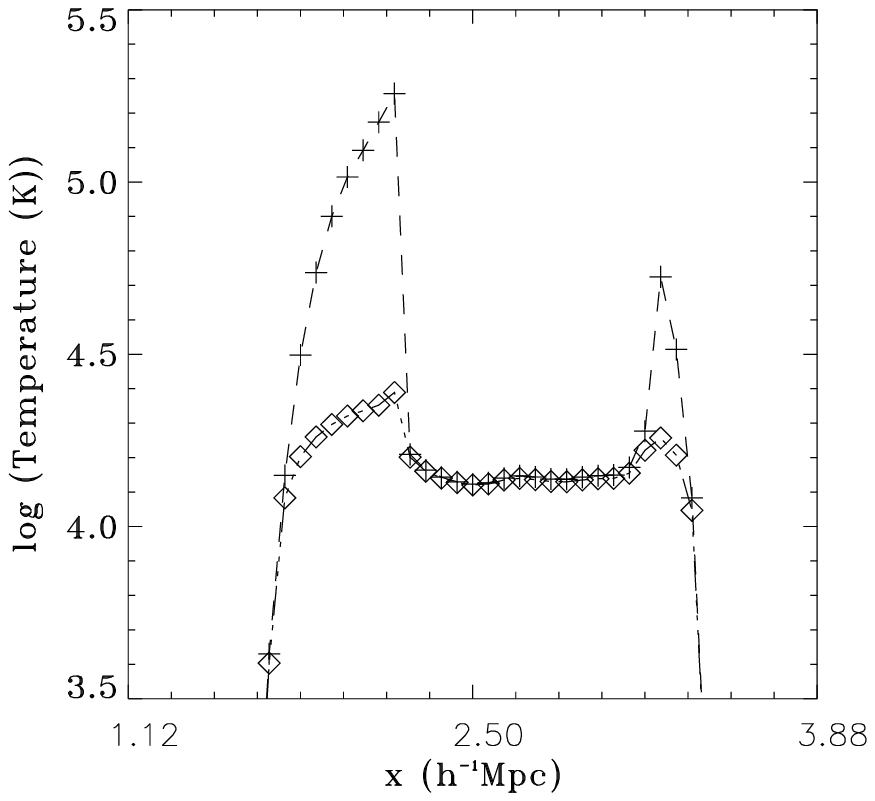}\\
\includegraphics[height=5.7cm]{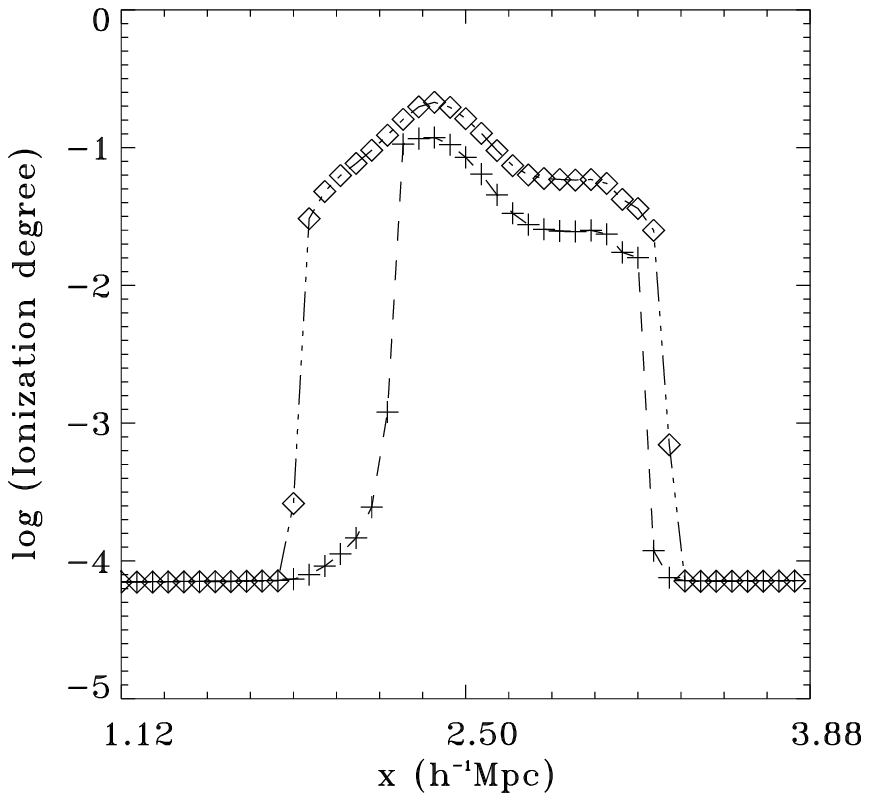}\\
\caption{Temperature and ionization degree profiles along the line of
  sight marked by the dashed line in Fig.~\ref{carteTS3T}: gas
  temperature in the $S_{1\mathrm T}$ simulation (top panel), heavy
  particle temperature (middle panel, plus sign), electron temperature
  (middle panel, diamond sign), ionization degree in the $S_{3\mathrm
  T}$ simulation (bottom panel, plus sign), ionization degree in the
  $S_{1\mathrm T}$ simulation (bottom panel, diamond sign).}
\label{profiles}
\end{center}
\end{figure}

Figure~\ref{carteTS3T} illustrates departure from equilibrium in
structures much less massive than galaxy clusters and at high
redshift. Temperature distributions for a typical region show
significant differences between the simulation with non-equipartition
processes and the simulation in which equipartition is forced between
species: heavy particle temperature in the $S_{3\mathrm T}$ simulation
is larger than plasma temperature in the $S_{1\mathrm T}$
simulation. These differences are located in the outer regions of
bound structures and are in the range $10^4$--$10^6$ K.

Figure~\ref{profiles} displays the temperature profiles along a line
of sight through one of the structures shown on the temperature
distributions in Fig.~\ref{carteTS3T} (the line of sight is marked by
the dashed line). The outer parts of the profiles show that, in the
$S_{3\mathrm T}$ simulation, the heavy particle temperature rises up
to $10^5$ K whereas the electron temperature is around $10^4$ K. In
the same regions but for the $S_{1\mathrm T}$ simulation the plasma
temperature is at a few times $10^4$ K. Then in the range of
temperature $10^4$--$10^5$ K, the IGM in the $S_{3\mathrm T}$
simulation is significantly warmer than in the $S_{1\mathrm T}$
simulation.

Figure~\ref{profiles} plots also the ionization degree profiles along
the same line of sight. The ionization degree is defined by
$x=n_\mathrm e/(n_\mathrm H+n_\mathrm{He})$. These profiles point out
that in the regions of temperature differences (clearly seen on the
left side of the profiles), the plasma is weakly ionized in the
$S_{3\mathrm T}$ simulation whereas it is partially ionized in the
$S_{1\mathrm T}$ simulation. The former being two orders of magnitude
lower than the latter. \\

\begin{figure*}
\begin{center}
\includegraphics[height=9cm]{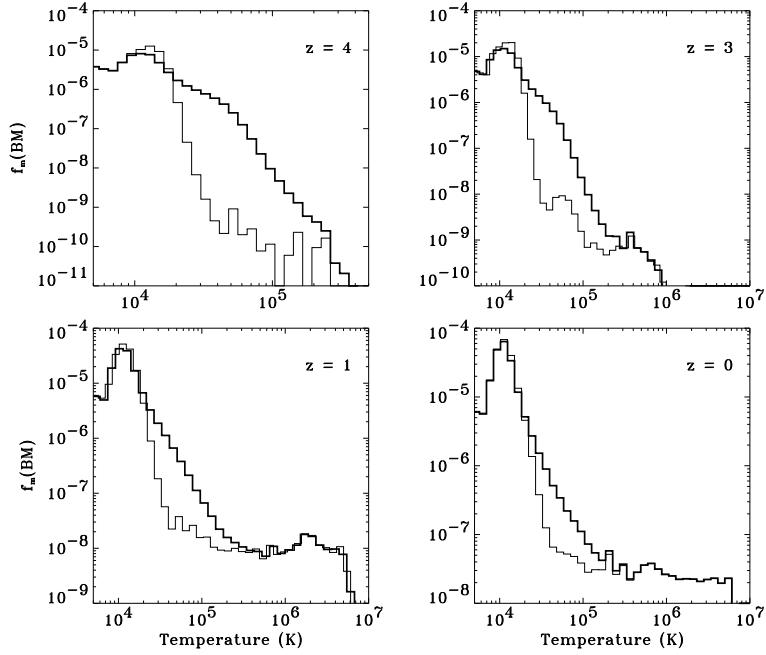}
\caption{Baryonic mass fraction per interval of electron
temperature (thin line) and per interval of heavy particle
temperature (thick line) at different redshifts in the $S_{3\mathrm T}$
simulation.}
\label{dist_lcdm}
\end{center}
\end{figure*}

We now quantitatively estimate the mass fraction of that warmer
plasma. First of all the baryonic mass fraction per interval of heavy
particle temperature and per interval of electron temperature is
computed in the $S_{3\mathrm T}$ simulation at different
redshifts. Figure \ref{dist_lcdm} shows that the baryonic mass
fraction with heavy particle temperature between $10^4$--$2.10^5$ K is
larger than the baryonic mass fraction with electron temperature in
the same range.  That warm out of equilibrium plasma exists at any
redshift but its mass fraction decreases with redshift.

\begin{figure*}
\begin{center}
\includegraphics[height=5.5cm]{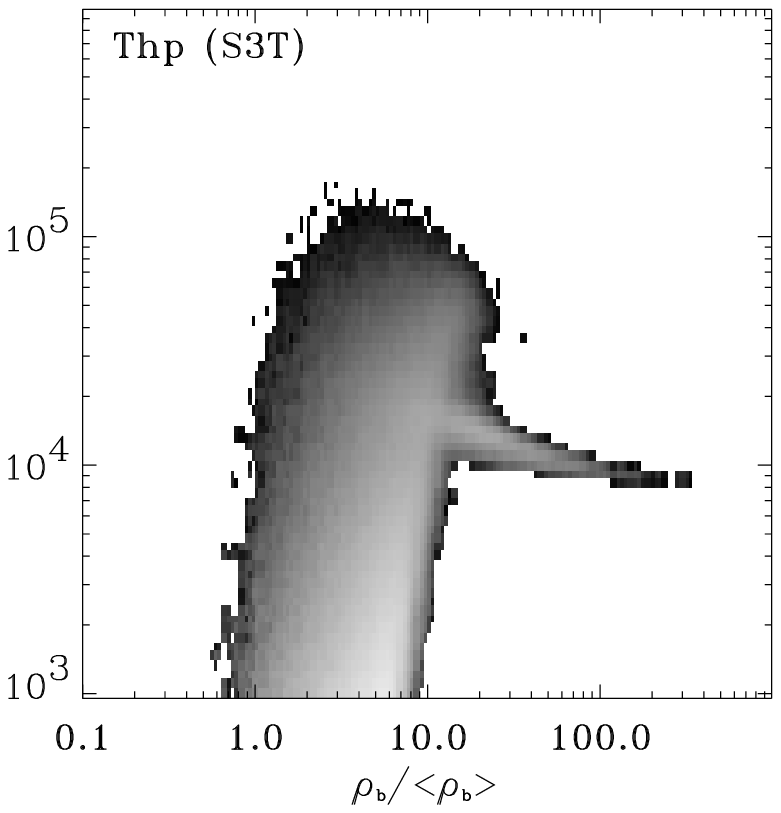}
\includegraphics[height=5.5cm]{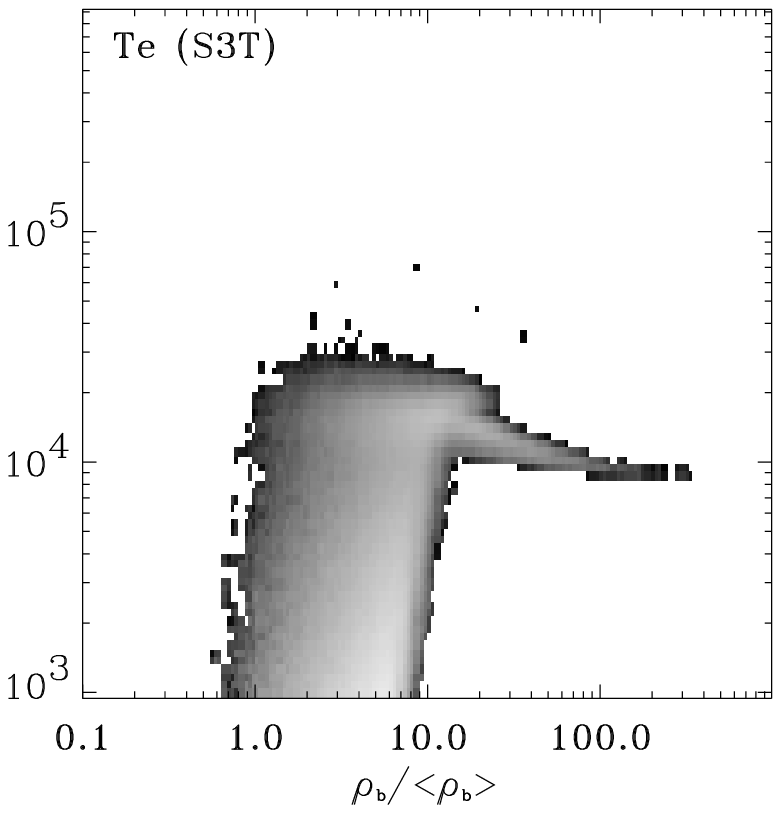}
\includegraphics[height=5.5cm]{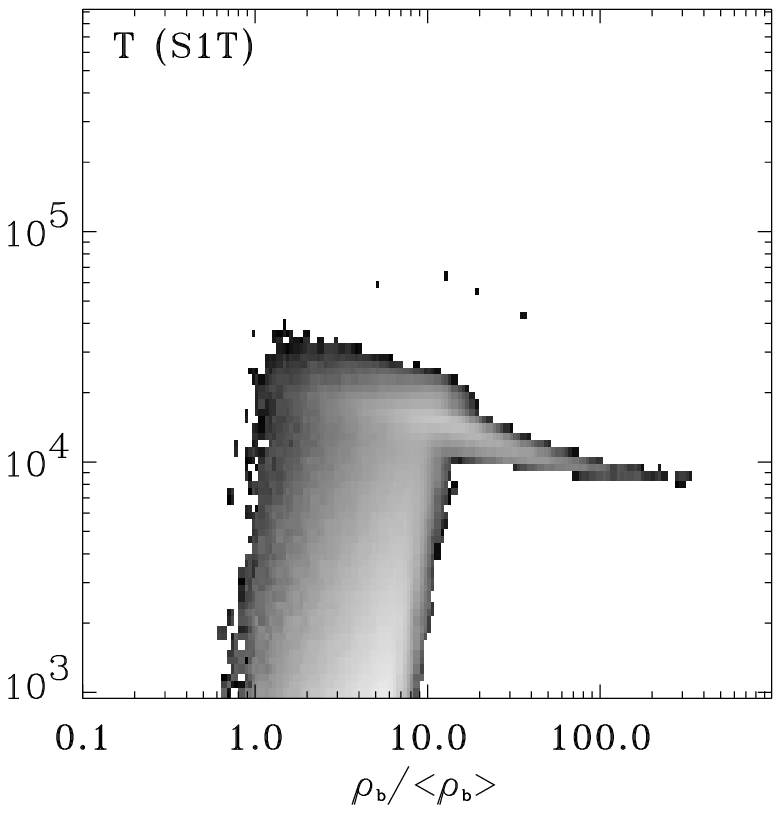}
\caption{Baryonic mass fraction at $z=5$ per interval of baryonic
  density contrast and per interval of heavy particle temperature
  (left panel), per interval of electron temperature (middle panel) in
  the $S_{3\mathrm T}$ simulation, per interval of gas temperature in
  the $S_{1\mathrm T}$ simulation (right panel), light grey is high
  mass fractions and temperature is shown on the ordinate axis.}
\label{iso_lcdm}
\end{center}
\end{figure*}

To inquire about the density of the warm plasma, we plot in Fig.
\ref{iso_lcdm} isocontours of the baryonic mass fraction per interval
of temperature and per interval of baryonic density contrast
$\delta_\mathrm b=\rho_\mathrm b/ \langle \rho_\mathrm b \rangle$. The
diagram on the left, as a function of the heavy particle temperature,
shows a region with temperatures in the range $2.10^4$--$5.10^5$ K and
baryonic density contrasts around 10 or less. The middle plot shows
that electron temperature in this region is a few times $10^4$ K. The
point here is that this warm plasma region is not found in the
$S_{1\mathrm T}$ simulation where the plasma is rather cold (right
panel). Differences between the two simulations are then identified by
a warm temperature plasma with relatively low baryonic density
contrast.

\begin{figure}
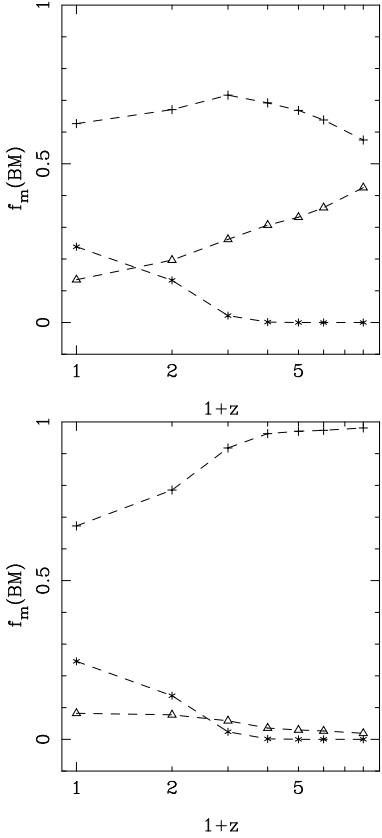

\begin{center}
\includegraphics[height=5cm, angle=-90]{fig6a.ps}
\includegraphics[height=5cm, angle=-90]{fig6b.ps}
\caption{Evolution with redshift of the baryonic mass fraction
  computed for different ranges of temperature (normalized to the
  total baryonic mass with a temperature higher than $9.10^3$ K):
  $9.10^3 \leq T < 2.10^4$ K (plus sign), $2.10^4 \leq T < 5.10^5$ K
  (triangle), $T \geq 5.10^5$ K (star), in the $S_{3\mathrm T}$ (upper
  panel) and $S_{1\mathrm T}$ (lower panel) simulations.}
\label{glob_lcdm}
\end{center}
\end{figure}

Finally we estimate the evolution with redshift of the different
phases of the plasma in the $S_{3\mathrm T}$ simulation and compare
with the $S_{1\mathrm T}$ simulation. We define a bulk plasma
temperature in the former simulation as: \begin{equation} \label{Tg}
T_\mathrm g = \frac{(s_\mathrm e+s_\mathrm i+s_\mathrm n)V_\mathrm
s^{-2/3}}{\frac{3}{2}kN_\mathrm{tot}} \equiv \frac{n_\mathrm
eT_\mathrm e+n_\mathrm iT_\mathrm i+n_\mathrm nT_\mathrm n}{n_\mathrm
e+n_\mathrm i+n_\mathrm n} \end{equation} where $N_\mathrm{tot}$ is
the total number of particles. The evolution of the baryonic mass
fraction is computed for different ranges of temperature, $T_\mathrm
g$ in $S_{3\mathrm T}$ and $T$ in $S_{1\mathrm T}$. Mass fractions are
normalized to the total baryonic mass with a temperature higher than
$9.10^3$ K. Three ranges of temperatures are considered: $9.10^3 \leq
T < 2.10^4$ K, $2.10^4 \leq T < 5.10^5$ K and $T \geq 5.10^5$ K.
Figure~\ref{glob_lcdm} shows that, between $z=7$ and $z=3$, about
$30$--$40\%$ of the plasma is warm ($2.10^4 \leq T < 5.10^5$ K) in the
$S_{3\mathrm T}$ simulation whereas this phase represents only less
than $5\%$ in the $S_{1\mathrm T}$ simulation. We note that, in
particular at high redshift, the plasma that is warm in the
$S_{3\mathrm T}$ simulation is cold ($9.10^3 \leq T < 2.10^4$ K) in
the $S_{1\mathrm T}$ simulation, and that the cold baryonic mass
fraction is much lower in the $S_{3\mathrm T}$ simulation (this point
will be addressed in the final section).

\subsection{Influence of the viscous dissipation function}
\label{Qv}

Before deducing any cosmological implications from the thermodynamic
and chemical differences between the two simulations as described in
the previous section, we have to check that our results are not due to
numerical effects. In fact the pseudo-entropy equation
(eq. \ref{DsDt}) computes the contribution to the energy of shocks and
gravitational compression and includes a numerical heating term given
by the artificial viscosity. Is this term responsible for departure
from equilibrium seen in the $S_{3\mathrm T}$ simulation? To
invalidate this hypothesis, we run the same simulations except that
now the constants $C_1$ and $C_2$ in the term $Q_\mathrm v$ (eq.
\ref{Qvterm}) are reduced by a factor 10.

\begin{figure*}
\begin{center}
\includegraphics[height=6cm]{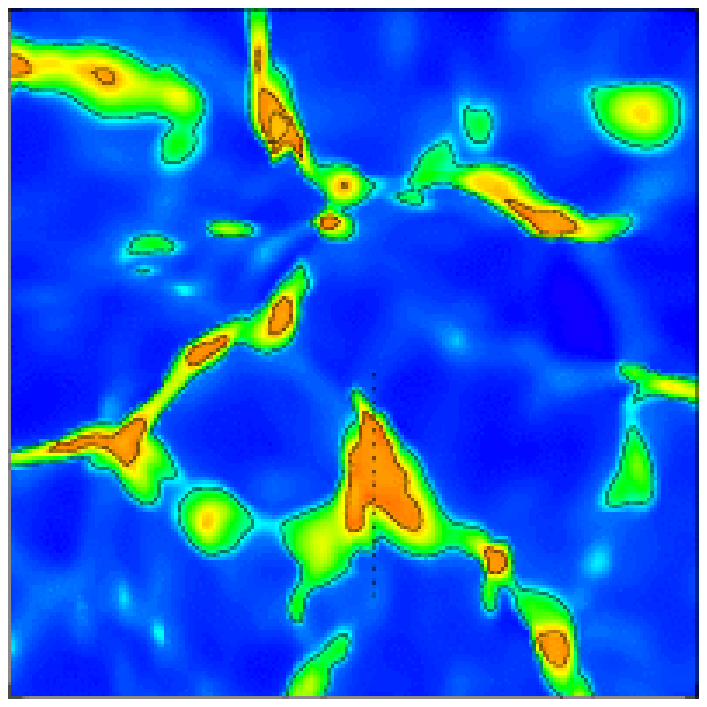}
\includegraphics[height=6cm]{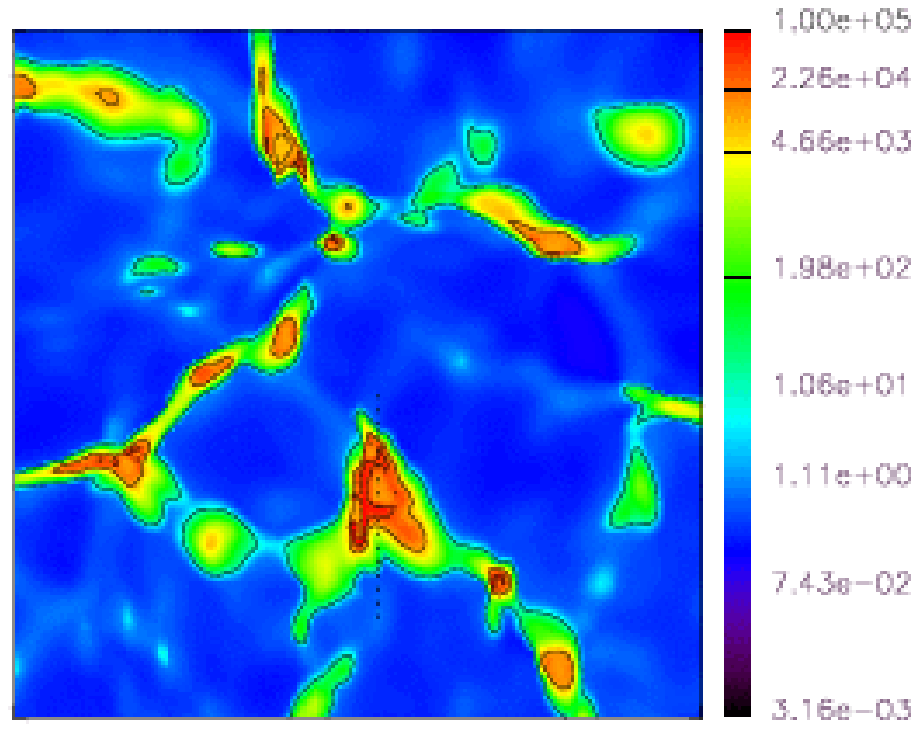}
\includegraphics[height=6cm]{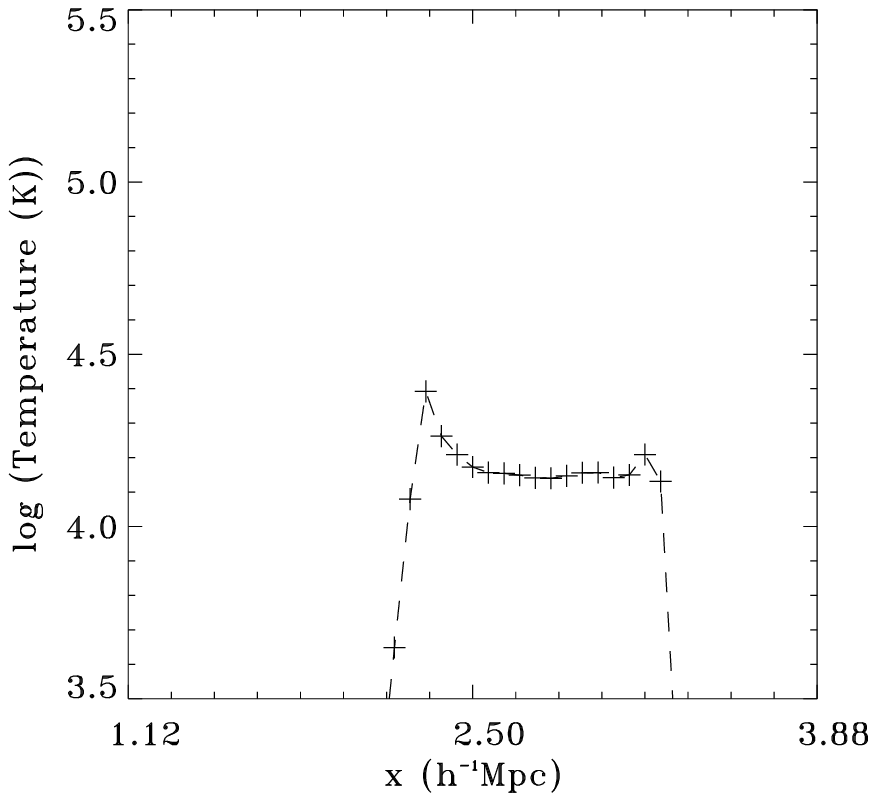}
\includegraphics[height=6cm]{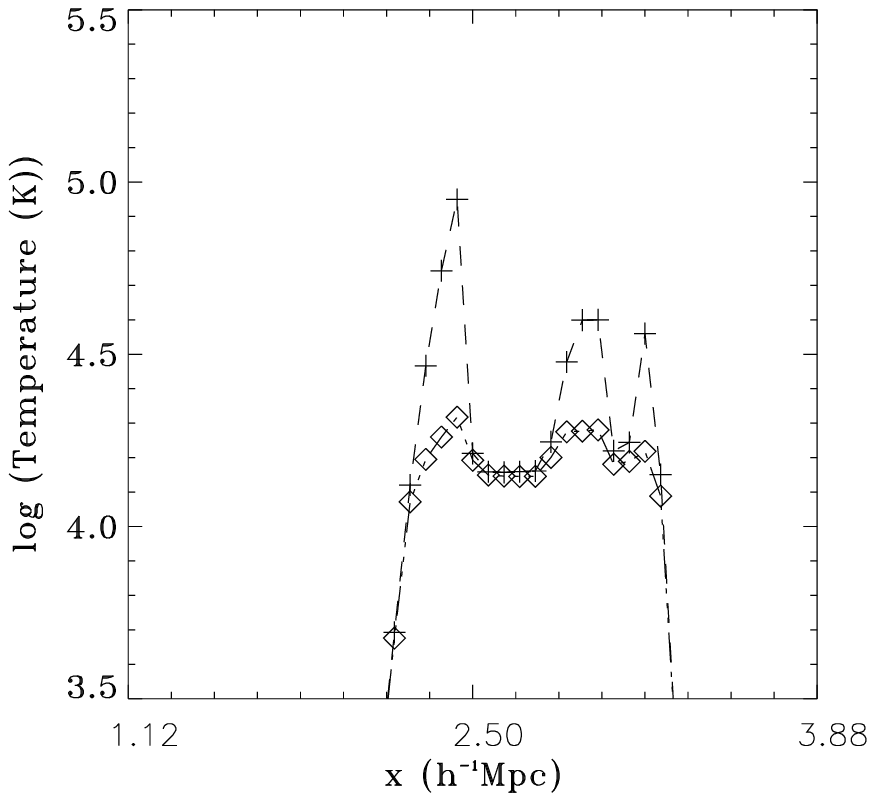}
\caption{Influence of the viscous dissipation function: the
  $S_{3\mathrm T}$ and the $S_{1\mathrm T}$ simulations are now
  computed with the $Q_\mathrm v$ term (eq. \ref{Qvterm}) reduced by
  10. Upper panels: isocontours of gas temperature in the $S_{1\mathrm
  T}$ simulation (left panel) and of heavy particle temperature in the
  $S_{3\mathrm T}$ simulation (right panel) for the same field as on
  Fig.~\ref{carteTS3T} (isocontours at $10., 10^4, 4.10^4$ K). Bottom
  panels: temperature profiles along the line of sight shown on top
  panels by the dashed line: gas temperature in the $S_{1\mathrm T}$
  simulation (left panel), heavy particle temperature (right panel,
  plus sign), electron temperature (right panel, diamond sign).}
\label{carteTS3T_Qv}
\end{center}
\end{figure*}

Figure~\ref{carteTS3T_Qv} plots the same temperature field as in
Fig.~\ref{carteTS3T} for heavy particles in the $S_{3\mathrm T}$
simulation and for plasma temperature in the $S_{1\mathrm T}$
simulation. The temperature profiles are also displayed along the same
line of sight as before. Temperature distribution in the $S_{1\mathrm
T}$ simulation shows that, due to the reduction of the shock
contribution, the amount of energy created by shocks is lower
resulting in a too low pressure to support gravitational compression.
The spatial extension of structures are then globally smaller than in
Fig.~\ref{carteTS3T}. \\

Nonetheless, comparing the two simulations reveals that the
$S_{3\mathrm T}$ simulation shows also a substantial fraction of the
IGM warmer than in the $S_{1\mathrm T}$ simulation and regions which
depart from equilibrium in the outer regions of bound structures are
spatially larger than the region over which the artificial viscosity
is applied.  We can conclude that the heating of heavy particles is
not dominated by numerical effects. Moreover as will be pointed out in
section \ref{highz}, increasing the resolution but keeping the same
value for the viscous term used in section \ref{low_strt}, leads also
to departure from equilibrium. The thermodynamic differences between
the $S_{3\mathrm T}$ and $S_{1\mathrm T}$ simulations are then of
physical origin and examined in the next section.

\section{Analytical interpretation}
\label{discuss}

In the outer regions of not too dense structures, the IGM is out of
equilibrium for temperatures in the range $10^4$--$10^6$ K and is
weakly ionized (Fig.~\ref{carteTS3T} and \ref{profiles}). Contrary to
the outer regions of galaxy clusters in which the IGM is totally
ionized and with sufficiently high density to give a relaxation
timescale of the order of the Hubble time (see eq. \ref{teion2}),
relaxation timescales between ions and electrons and between neutrals
and electrons are now very short compared to the Hubble time :
\begin{equation}
\tau_\mathrm{e,ion} \ll \tau_\mathrm{e,neut} \ll \tau_\mathrm H.
\end{equation}
We take into account interactions between neutrals and electrons since
the plasma is weakly ionized.

What are the mechanisms which control departure from equilibrium and
is the relaxation timescale long enough to allow physical effects on
cosmological time? This section provides answers to these questions
based on a semi-analytical model.\\

We consider here the relaxation of an out of equilibrium hydrogen
plasma having just undergone gravitational compression. Relaxation
processes are cooling and non-equipartition processes. We neglect
other processes such as the expansion of the universe or compression
due to the accretion of matter or merger events. The plasma is
composed of hydrogen nuclei (neutrals and ionized particles) and
electrons with number densities $n_\mathrm H \equiv n_\mathrm{HP} =
n_{\mathrm H^0} +n_{\mathrm H^+}$ and $n_\mathrm e$,
respectively. Densities can be expressed with the ionization degree
$x$ :
\begin{eqnarray}
\label{ne}
n_\mathrm e &=& xn_\mathrm H \equiv n_{\mathrm H^+} \ \textrm{and} \
n_{\mathrm H^0} = n_\mathrm H(1-x).
\end{eqnarray} 
Initial conditions are suggested by the results of the simulations:
temperatures of heavy particles and electrons are respectively,
$T_\mathrm{HP}=10^5$ K and $T_\mathrm e=2.10^4$ K, hydrogen nuclei
density is $n_\mathrm H=5.10^{-5} \ \textrm{cm}^{-3}$ and the
ionization degree is $x=10^{-4}$. \\

\begin{figure*}
\begin{center}
\includegraphics[height=10cm]{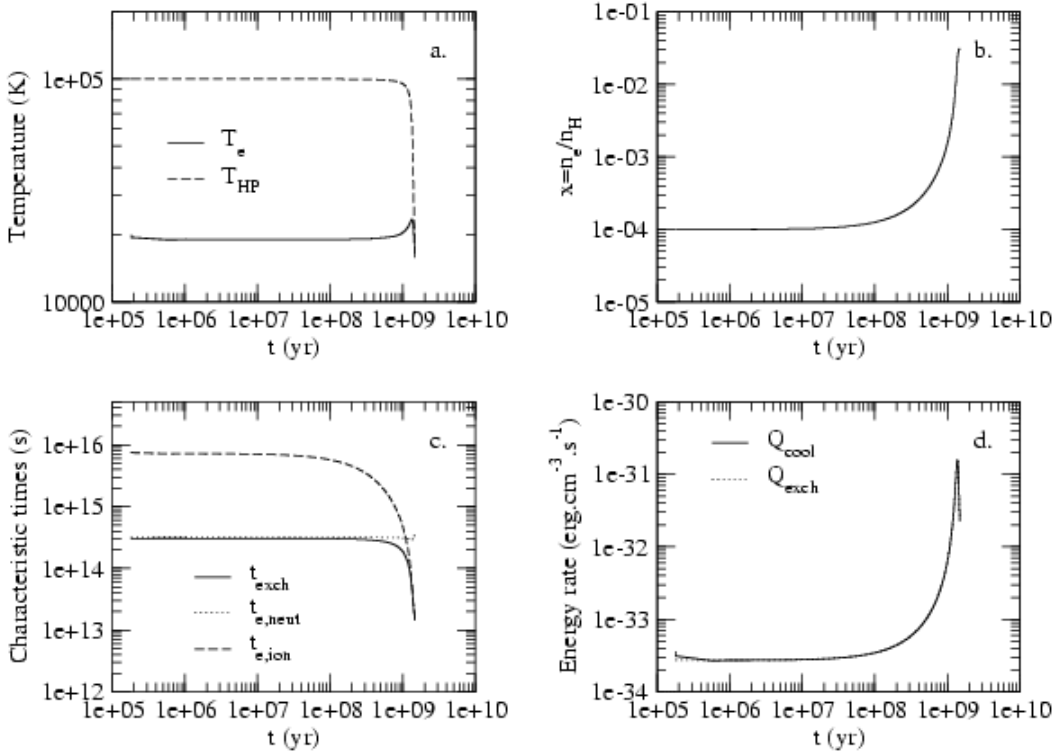}
\caption{Temporal evolution of heavy particle and electron
  temperatures (a), ionization degree (b), characteristic times (c),
  and energy exchange and cooling rates (d) using equations
  (\ref{dTLdt2}), (\ref{dTedt2}) and (\ref{dnedt}).}
\label{resu}
\end{center}
\end{figure*}

We solve an equation of energy balance for heavy particles and for
electrons: 
\begin{eqnarray}
\label{dhLdt}
\left( \frac{\mathrm{d}e}{\mathrm{d}t} \right)_\mathrm{HP} &=&
-\frac{1}{\rho}Q_\mathrm{exch}, \\
\label{dhedt}
\left( \frac{\mathrm{d}e}{\mathrm{d}t} \right)_\mathrm e &=&
+\frac{1}{\rho}Q_\mathrm{exch} -\frac{1}{\rho}Q_\mathrm{cool},
\end{eqnarray} 
where the right hand terms represent the change in the specific
internal energy for the two fluids. $Q_\mathrm{cool}$ is the rate of
energy lost by the electrons (per unit of volume and per unit of
time):
\begin{equation}
\label{qcool}
Q_\mathrm{cool} = \Lambda_\mathrm{exc}(\mathrm H) +
\Lambda_\mathrm{ion}(\mathrm H) + \Lambda_\mathrm{rec}(\mathrm H) +
\Lambda_\mathrm{brem}(\mathrm H).
\end{equation}
The cooling rates associated with each process, collisional
excitation, collisional ionization, recombination, bremsstrahlung, are
given in Table~\ref{coolrates}. $Q_\mathrm{exch}$ is the rate of
energy exchange between electrons and heavy particles (per unit of
volume and per unit of time):
\begin{equation}
Q_\mathrm{exch} = \frac{3}{2} kn_\mathrm e
\frac{(T_\mathrm{HP}-T_\mathrm e)}{\tau_\mathrm{exch}},
\end{equation}
where $\tau_\mathrm{exch}$ includes two terms due to two interaction
mechanisms, that between charged particles and between neutral and
charged particles:
\begin{equation}
\frac{1}{\tau_\mathrm{exch}}=\frac{1}{\tau_\mathrm{e,ion}} +
\frac{1}{\tau_\mathrm{e,neut}}.
\label{texch}
\end{equation}
We consider isochoric transformations, $\mathrm{d}\rho /
\mathrm{d}t=0$, then, since $\rho \approx m_\mathrm p n_\mathrm H$
with $m_\mathrm p$ the atomic mass, $\mathrm{d}n_\mathrm H /
\mathrm{d}t=0$ and equations (\ref{dhLdt}) and (\ref{dhedt}) become:
\begin{eqnarray}
\label{dTLdt2}
\frac{3}{2} kn_\mathrm H \frac{\mathrm{d}T_\mathrm{HP}}{\mathrm{d}t}
&=& -Q_\mathrm{exch},\\
\label{dTedt2}
\frac{3}{2} kn_\mathrm e \frac{\mathrm{d}T_\mathrm e}{\mathrm{d}t} &=&
+Q_\mathrm{exch} -Q_\mathrm{cool} -\frac{3}{2} kT_\mathrm e n_\mathrm
H \frac{\mathrm{d}x}{\mathrm{d}t}, \end{eqnarray} where $x$ is the
ionization degree. The last term of the second equation is computed
from the evolution equation of the electron density:
\begin{equation}
\label{dnedt}
\frac{\mathrm{d}n_\mathrm e}{\mathrm{d}t} = n_\mathrm
H\frac{\mathrm{d}x}{\mathrm{d}t} = \beta_{\mathrm H^0} n_\mathrm
en_{\mathrm H^0} - \alpha_{\mathrm H^+} n_\mathrm en_{\mathrm H^+},
\end{equation}
where $\beta_{\mathrm H^0}$ and $\alpha_{\mathrm H^+}$ are
respectively the ionization and recombination rates for hydrogen (see
Table~\ref{alphabetarates}).  We compute the evolution of $T_\mathrm
e$, $T_\mathrm{HP}$ and $x$ with respect to time using equations
(\ref{dTLdt2}), (\ref{dTedt2}) and (\ref{dnedt}).  \\

Figure~\ref{resu}a shows that departure from equilibrium is maintained
on a timescale of gigayears, and that during this time the ionization
degree does not evolve (Fig.~\ref{resu}b). It is interesting to note
that just before the equilibration between the two temperatures, the
electron temperature slightly increases because of the energy gained
in elastic processes\footnote{Similar evolutions were computed in a
sophisticated analysis by \cite{Fadeyev}, of the structure of
radiative shock waves propagating in stellar envelopes.}. Figure
\ref{resu}c presents the evolution of the different characteristic
times: the energy exchange timescale between electrons and ions (eq.
\ref{teion}), the energy exchange timescale between electrons and
neutrals (eq. \ref{teneut}), and the resultant $\tau_\mathrm{exch}$
due to both mechanisms (eq. \ref{texch}). It is interesting to see
that in the relaxation region and since the plasma is weakly ionized,
elastic processes are dominated by neutral-electron interactions and
$\tau_\mathrm{exch} \approx \tau_\mathrm{e,neut}$. Equilibration is
recovered once plasma is getting partially ionized and during this
short period, $\tau_\mathrm{exch} \approx \tau_\mathrm{e,ion}$. Once
the ionization degree increases, $\tau_\mathrm{exch}$ decreases
drastically and equipartition is established. \\

\begin{figure}
\begin{center}
\includegraphics[width=6.5cm, angle=-90]{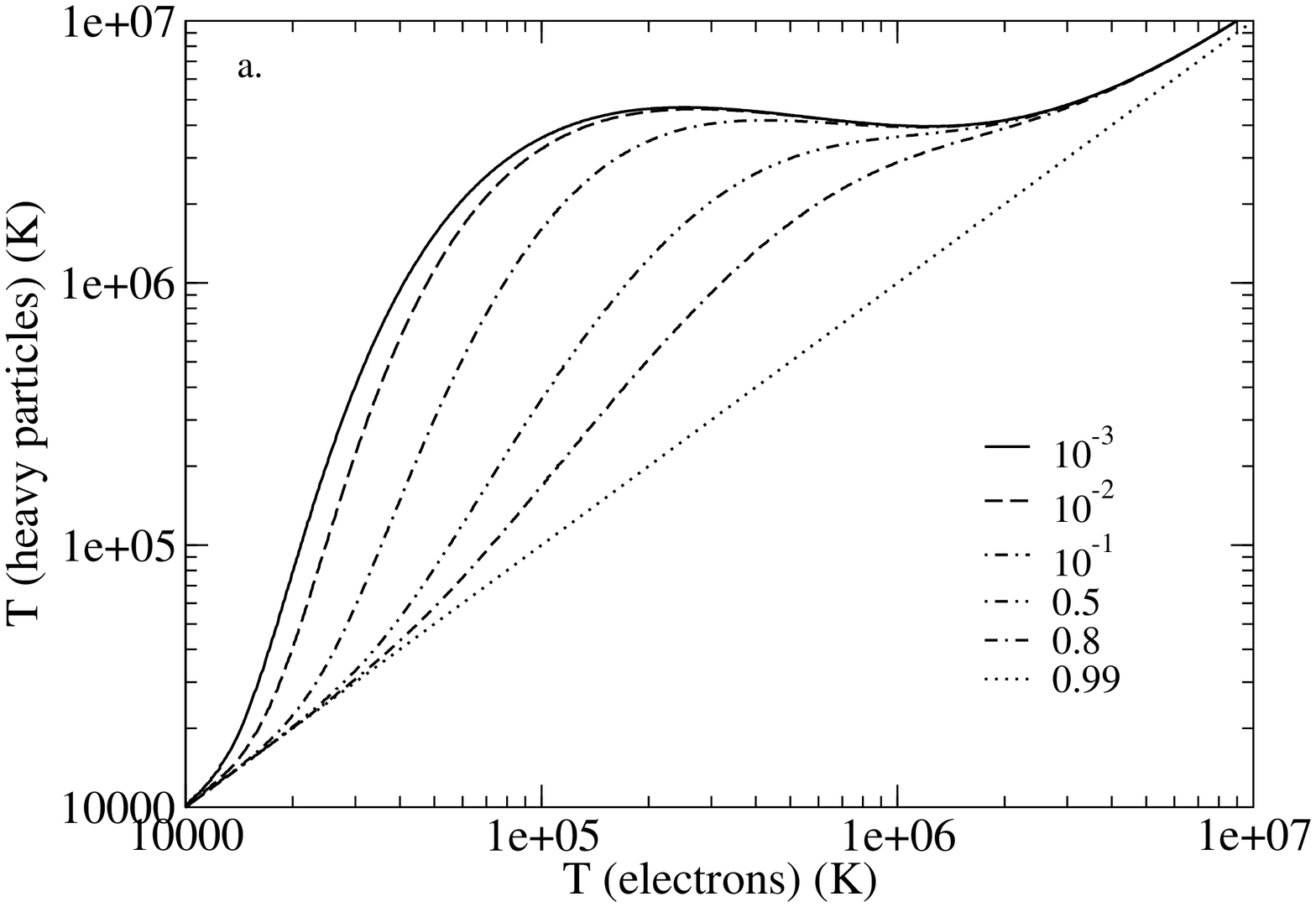}
\includegraphics[width=6.5cm, angle=-90]{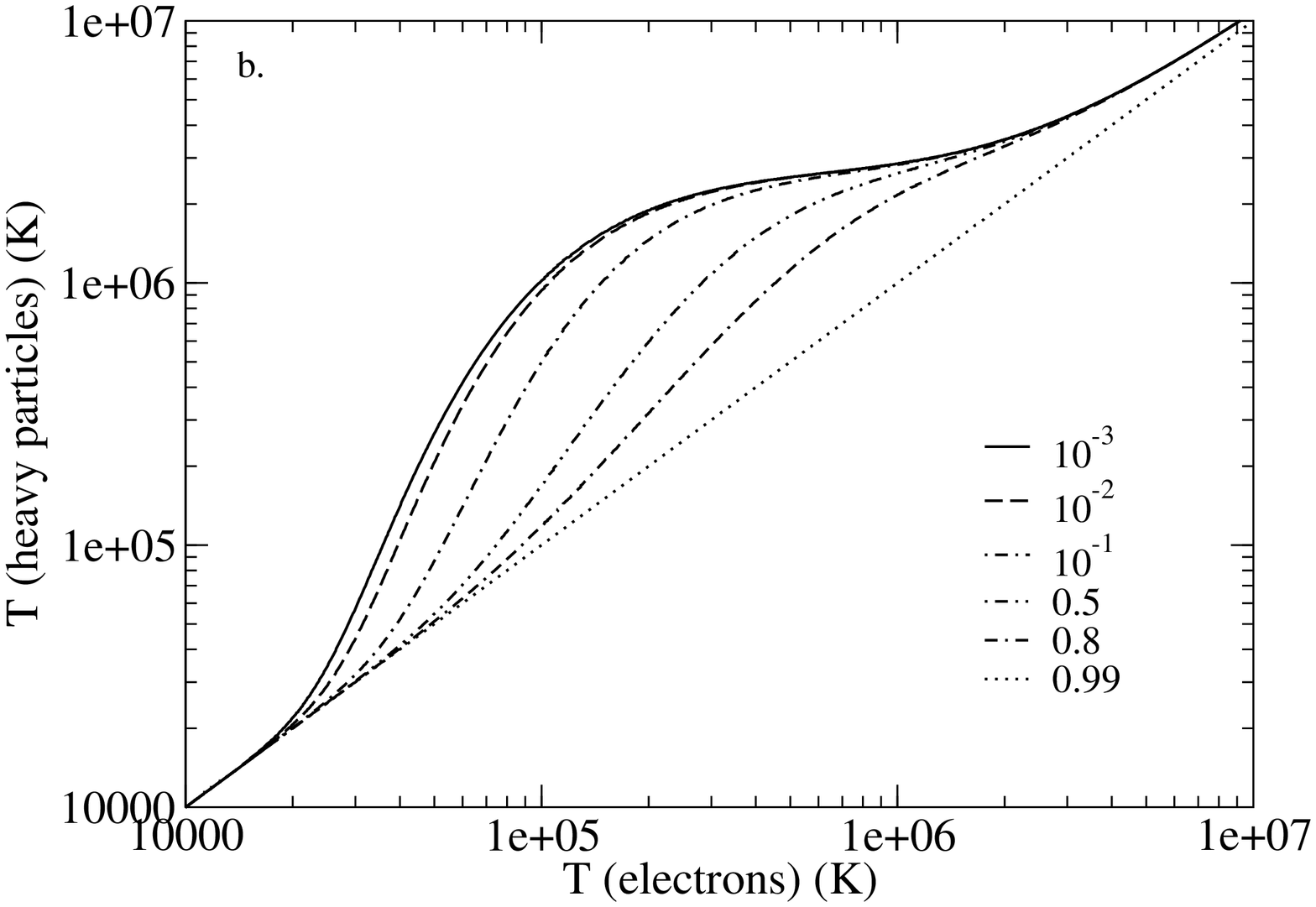}
\caption{Heavy particle temperature versus electron temperature for
  different values of the ionization degree computed by equation
  (\ref{relTLTe}) with (a) and without (b) collisional excitation
  cooling term in eq. (\ref{qcool}).}
\label{TLTe}
\end{center}
\end{figure}

Moreover, Fig.~\ref{resu}d shows that the energy exchange rate between
heavy particles and electrons is approximately equal to the cooling
rate. This means that there is no change in the electronic internal
energy (eq. \ref{dhedt}). By writing $Q_\mathrm{exch} \approx
Q_\mathrm{cool}$ we then derive a relation between $T_\mathrm{HP}$ and
$T_\mathrm e$ such as:
\begin{equation}
\label{relTLTe}
T_\mathrm{HP} = T_\mathrm e + (\frac{3}{2}\frac{kn_\mathrm e}{\tau_\mathrm{exch}})^{-1}Q_\mathrm{cool}.
\end{equation}
Since $Q_\mathrm{cool} = f(T_\mathrm e, n_\mathrm e, n_{\mathrm H^0},
n_{\mathrm H^+})$ this relation is only parameterized by the ionization
degree\footnote{This approach was used by \cite{Petschek} to determine
the ionization rate of a gas of argon in a shock tube.}. In
Fig.~\ref{TLTe}a we plot $T_\mathrm{HP}$ as evaluated in equation
(\ref{relTLTe}), with the cooling term given by equation
(\ref{qcool}), for different values of the ionization degree.  In a
narrow electronic temperature range, $10^4$--$3.10^4$ K, the heavy
particle temperature extends in a much larger range, between $10^4$
and $5.10^5$ K. It is interesting to note that the higher the value of
the ionization degree, the smaller is the difference between the two
temperatures. For very low ionization degree the relation between
$T_\mathrm{HP}$ and $T_\mathrm e$ does not depend on $x$ since
$\tau_\mathrm{e,neut} \propto (n_{\mathrm H^0} T_\mathrm e^{1/2})^{-1}$,
and Fig.~\ref{resu}c shows that departure from equilibrium in weakly
ionized plasma is controlled by neutral-electron mechanisms.

In this range of electronic temperatures, cooling via bremsstrahlung
processes is not efficient, and neither is cooling due to
recombination processes (too low density). Then only terms due to
collisional excitation and collisional ionization processes are
effective in equation (\ref{qcool}). Figure~\ref{TLTe}b illustrates
the relation between $T_\mathrm{HP}$ and $T_\mathrm e$ when equation
(\ref{relTLTe}) is solved by neglecting the collisional excitation
term. We note that even in that case, departure from equilibrium is
also seen, but now for higher electron temperatures. \\

\begin{figure}
\begin{center}
\includegraphics[height=9cm, angle=-90]{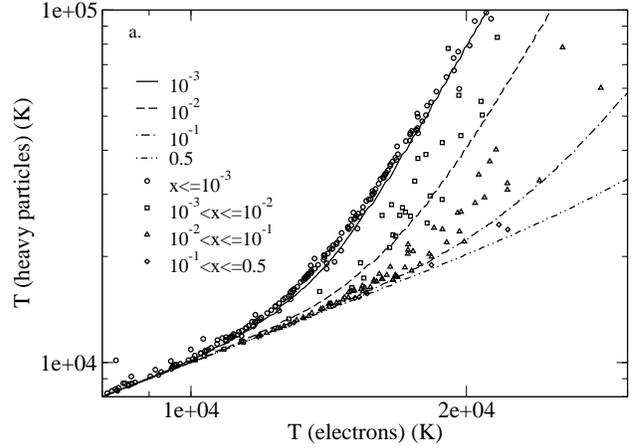}
\caption{Heavy particle temperature versus electron temperature at
  $z=3$ in 300 cells randomly chosen in the $S_{3\mathrm T}$
  simulation; each symbol represents the ionization degree in each
  cell and the curves are computed in the same way as in
  Fig.~\ref{TLTe}a.}
\label{TLTe_cell}
\end{center}
\end{figure}

Then for a given ionization degree and electron temperature slightly
higher than $10^4$ K, the energy exchange rate between the heavy
particles and the electrons compensates for the energy lost by the
electron gas due to inelastic processes. We now return to the
$S_{3\mathrm T}$ simulation and confirm this scenario by plotting
temperature values for heavy particles and electrons in each cell of
the computational volume. Each symbol in Fig.~\ref{TLTe_cell}
corresponds to the range of ionization degree of each cell of the
simulation. We overplot relation (\ref{relTLTe}) for different
ionization degrees. For instance squares indicate cells with
ionization degree between $10^{-3}<x \le 10^{-2}$, and we see that
these cells are in the region delimited by the curves $x=10^{-3}$ and
$x=10^{-2}$. The agreement for each range of ionization validates the
condition $Q_\mathrm{exch} \approx Q_\mathrm{cool}$. These results
confirm our understanding of the physical origin of the
non-equilibrium IGM in moderately dense structures, although numerical
simulations include a larger set of physical processes (as expansion
and accretion).\\

Departure from equilibrium results from the competition between
elastic processes in the weakly ionized IGM and cooling processes. It
is clear that the thermodynamic history of accreted gas in bound
structures is not the same in a simulation with non-equipartition
processes and in a simulation in which equipartition is forced. As the
plasma undergoes gravitational compression in not too dense
structures, heavy particles are heated much more than the electrons.
The electron temperature does not increase sufficiently to partially
ionize the medium (see also the evolution of temperature and
ionization degree in Fig.~\ref{resu}). Therefore the ionization degree
profiles in the outer regions in the $S_{3\mathrm T}$ simulation (Fig.
\ref{profiles}) show a weakly ionized plasma with $T_\mathrm e \sim
10^4$ K.  But in the $S_{1\mathrm T}$ simulation the electrons can be
heated up to $2.10^4$ K, and though the ionization temperature of
hydrogen is, with $I$ the ionization potential, $T=I/k\sim 1.6 \times
10^5$ K, a plasma at $T \simeq 2.10^4$ K is $\sim 90\%$ ionized
\citep{Shchekinov}. The ionization degree of a hydrogen plasma in
collisional ionization equilibrium is $\beta_{\mathrm
H^0}/(\beta_{\mathrm H^0}+\alpha_{\mathrm H^+})$. This fact explains
the two orders of magnitude difference between the ionization degree
profiles plotted for the $S_{3\mathrm T}$ and the $S_{1\mathrm T}$
simulations (at $1.5 \times 10^4$ K, the ionization rate for hydrogen
is $\sim 10^{-13}$ $\mathrm{s}^{-1} \ \mathrm{cm}^3$ whereas at $2.5
\times 10^4$ K the ionization rate is $\sim 10^{-11}$ $\mathrm{s}^{-1}
\ \mathrm{cm}^3$). \\

Before closing this section, we would like to mention an interesting
point.  In the center of structures such as in Fig.~\ref{carteTS3T},
the IGM is cold and partially ionized in the $S_{1\mathrm T}$
simulation (Fig.  \ref{profiles}). This suggests that in such not too
dense structures the IGM can be never heated to the so-called
virialized temperature and can be partially ionized and radiate its
energy just by being heated up to $3.10^4$ K.  This is supporting
evidence to the important question recently summarized by
\cite{Katz2002} concerning results from different numerical
simulations of large scale structure formation, showing that a
non-negligible fraction of the accreted gas can never reach the
virialized temperature \citep{Katz91, Kay, Fardal}.

\section{Cosmological implications}
\label{csq}

\begin{figure*}
\begin{center}
\includegraphics[height=6cm]{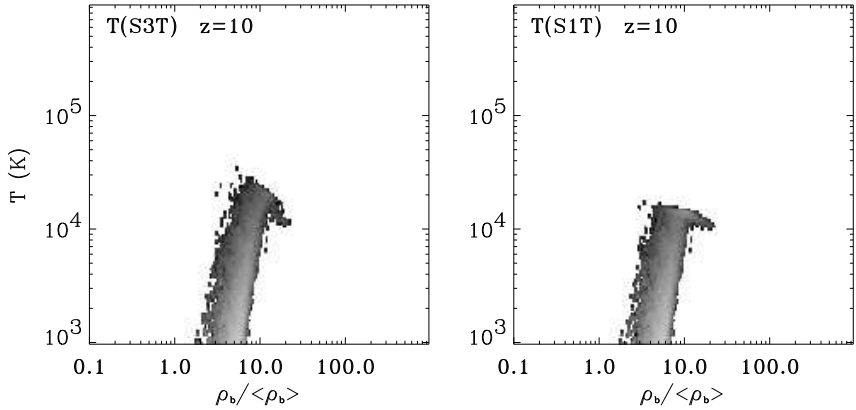}
\caption{Baryonic mass fraction at $z=10$ per interval of the baryonic
  density contrast and per interval of the bulk temperature in the
  $S_{3\mathrm T}$ simulation (left panel) and per interval of gas
  temperature in the $S_{1\mathrm T}$ simulation (right panel), both
  simulations including photoionization processes.}
\label{iso_fm1L16_phot_3t}
\end{center}
\end{figure*}

The introduction of non-equipartition processes alters, for not too
dense structures, the properties of plasma with temperature in the
range $10^{4}$--$10^{6}$ K. This plasma is expected to contribute to
galaxy formation. In this section we qualitatively analyze the
influence of non-equipartition processes on galaxy formation. We
discuss when and how this influence is dominant.

\subsection{At high redshift}
\label{highz}

The analytical interpretation shows that the weaker the plasma is
ionized the larger the departure from equilibrium. Figure~\ref{TLTe}
shows that the difference between heavy particle and electron
temperature (in the range $10^{4}$--$10^{6}$) is larger for lower
ionization degree and Fig.~\ref{resu} shows that decoupling is mainly
due to interaction mechanisms between neutrals and electrons. This
suggests that two conditions are required for a not too dense plasma
to be out of equilibrium. The first one is that the plasma accreted
into structures must be weakly ionized. The second one is that cooling
processes must involve atomic cooling (collisional excitation and/or
collisional ionization). This suggests that the major influence of the
non-equipartition processes occurs before the end of the reionization
epoch, i.e. at epochs when the cosmological plasma is not fully
ionized and when the cooling is dominated by atomic cooling.

\begin{figure}
\begin{center}
\includegraphics[height=7.5cm, angle=-90]{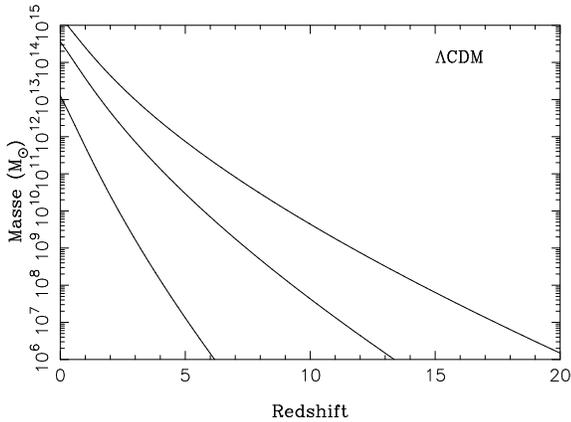}
\caption{Collapsing redshift and mass of $(1$--$2$--$3)\sigma$
  fluctuations (from left to right), computed by a spherical top-hat
  collapse model (eq. \ref{nulcdm}) in a $\Lambda$-cold dark matter
  universe.}
\label{collapse}
\end{center}
\end{figure}

To numerically confirm this point, we introduce photoionization
processes in the $S_{3\mathrm T}$ and the $S_{1\mathrm T}$ simulations
to see what happens after complete reionization of the universe. As
the analysis of the influence of photoionization processes is not the
aim of this paper, we do not give the details of the numerical aspects
of this implementation (see Alimi $\&$ Courty, in preparation).
Photo-ionization rates and heating rates are computed from the
evolution of the hydrogen and helium densities and from the spectrum
of the ultraviolet radiation background. Following \cite{Weinberg97}
the photoionizing background is assumed to have a redshift-dependent
evolution, with the bulk of the transition between $z=7$ and $z=6$ and
the peak between $z=3$ and $z=1$.  Moreover, as shown in the numerical
study by \cite{Gnedin2000}, we take into account a shallow evolution
of the ultraviolet background intensity before $z \sim 7$ and begin
the ionization processes at the redshift of $11.5$. The radiation is
considered as a spatially uniform field over the computational volume.
We compute the baryonic mass fraction per interval of baryonic density
contrast and per interval of temperature, which is the bulk
temperature for the $S_{3\mathrm T}$ simulation (eq. \ref{Tg}). Figure
\ref{iso_fm1L16_phot_3t} shows that at $z=10$ a fraction of the plasma
is warmer in the simulation with non-equipartition processes than in
the $S_{1\mathrm T}$ simulation. At $z=8$ no differences can be seen
between the two simulations. This suggests that the IGM is warmer in a
simulation with non-equipartition processes only before the universe
has been completely reionized. However to draw any firm conclusions
concerning the temperature differences between heavy particles and
electrons before this epoch, simulations including a more careful
treatment of how reionization proceeds should be performed, which is
currently out of reach.  \\

\begin{figure*}
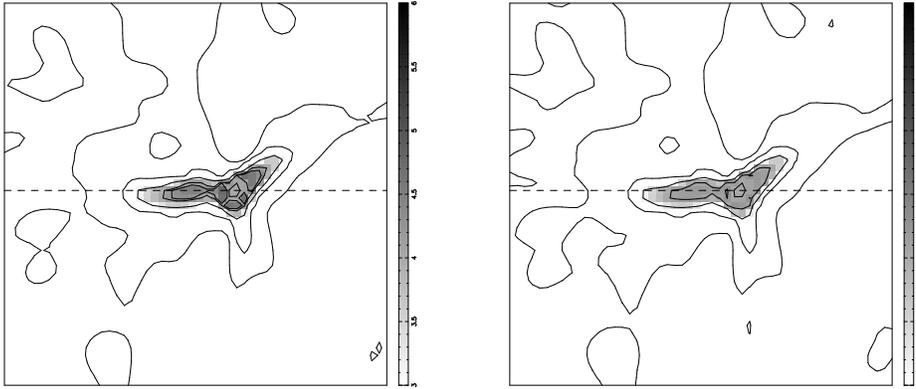

\begin{center}
  \includegraphics[height=5.5cm, angle=-90]{fig13a.ps}
  \hspace{1cm}
  \includegraphics[height=5.5cm, angle=-90]{fig13b.ps}
\caption{Heavy particle (left panel) and electron (right panel)
  temperature distributions at $z=12$ for the high resolution
  $S_{3\mathrm T}$ simulation with $N_\mathrm p=N_\mathrm g=384^3$
  (the slices are $2.166$ comoving $h^{-1} \textrm{Mpc}$ on a side and
  contour levels are: 2., 3., 10., $10^3$, $1.1 \times 10^4$, $1.7
  \times 10^4$, $2.5 \times 10^4$, $2.10^5$ K).}
\label{z12}
\end{center} 
\end{figure*} 

\begin{figure}
\begin{center}
\includegraphics[height=7.5cm, angle=-90]{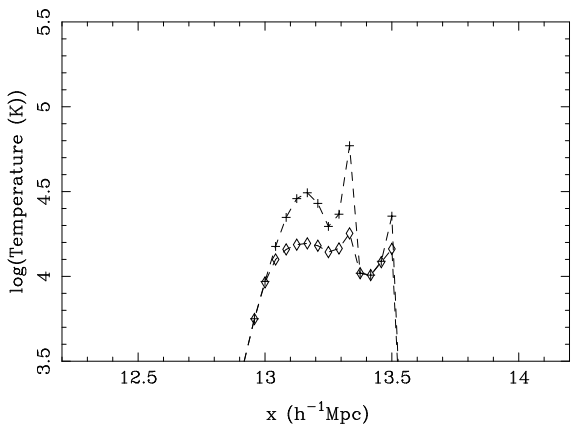}
\caption{Temperature profiles along a line of sight (dashed line in
  Fig.~\ref{z12}) in the structure at $z=12$ for the high resolution
  $S_{3\mathrm T}$ simulation: heavy particle temperature (plus sign),
  electron temperature (diamond sign).}
\label{profz12}
\end{center}
\end{figure}

As mentioned above, non-equipartition requires structures such that
cooling is dominated by atomic processes and the question is now to
examine whether such structures exist at high redshift. Figure
\ref{collapse} illustrates the mass and the collapsing redshift of a
$\nu$--$\sigma$ fluctuation computed in the top-hat collapse spherical
model and with cosmological parameters of our $\Lambda$ cold dark
matter model. Here, $\nu$ is the relative amplitude of fluctuation of
scale $M$ in units of the filtered dispersion $\sigma(M,z=0)$ at
$z=0$:
\begin{equation} 
\nu = \frac{\delta^\mathrm{coll}_\mathrm{lin}}{\sigma(M,z=0) D(z)},
\label{nulcdm}
\end{equation}
with $\delta^\mathrm{coll}_\mathrm{lin}$ the critical spherical
overdensity at $z=0$ and $D(z)$ the growth factor
\citep{Padmanabhan}. Before $z=10$, which is likely before the end of
the epoch of reionization, $3\sigma$ fluctuations have already
collapsed. These curves are plotted in a simplified way and we refer
to \cite{Hutchings} for recent studies. In a virialized approach,
structures with mass higher than a minimal mass of around $10^8$
M$_{\sun}$ are known to cool efficiently by atomic processes and the
so-called virialization temperature is larger than $10^4$ K. But as
numerical simulations do not make any assumptions about the
temperature of the IGM, we perform a high resolution simulation taking
non-equipartition processes into account to check if structures with
an out of equilibrium plasma are present at redshift higher than the
epoch of reionization. The number of dark matter particles and grid
cells is now equal to $384^3$ for the same computational box length,
allowing the initial density fluctuation spectrum to include smaller
scale modes.

\begin{figure}
\begin{center}
  \includegraphics[height=8cm, angle=-90]{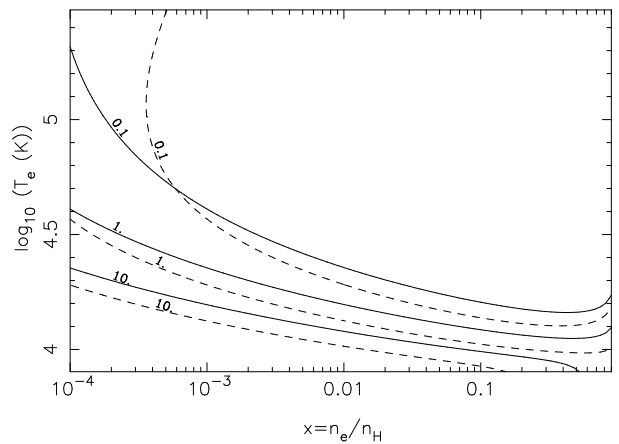}
\caption{Isocontours in a electron temperature-ionization
degree diagram of cooling timescales computed with equations (\ref{tcoolNE})
(dashed curves) and (\ref{tcoolE}) (solid curves) normalized to the
dynamical timescale (eq. \ref{tdyn}), the labels on each curves are the
ratios $t_\mathrm{cool}/t_\mathrm{dyn}$.}
\label{compt_tps3}
\end{center}
\end{figure}

\begin{figure*}
\begin{center}
\includegraphics[height=10cm]{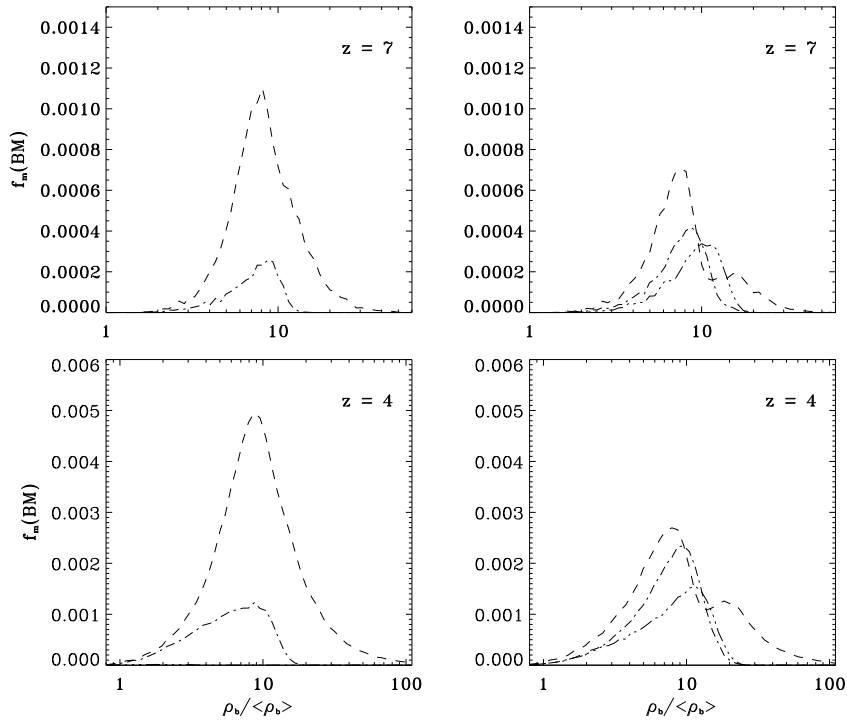}
\caption{Baryonic mass fraction per interval of the baryonic density
  contrast computed at two redshifts for different ranges of
  temperature: $8.10^3 \leq T < 1.7 \times 10^4$ K (dashed line), $1.7
  \times 10^4 \leq T < 3.10^4$ K (dot-dashed line), $3.10^4 \leq T <
  5.10^5$ K (triple dot-dashed line) in the $S_{3\mathrm T}$ (right
  panels) and the $S_{1\mathrm T}$ (left panels) simulations.}
\label{fmBM_BM}
\end{center}
\end{figure*}

This high resolution simulation is then able to catch bound structures
at very high redshift. An example of such a structure at $z=12$ is
illustrated in Fig.~\ref{z12} and temperature profiles along an
horizontal line of sight through the structure is plotted in Fig.
\ref{profz12}. Similar differences between the electron and heavy
particle temperatures can be seen in the outer regions of the
structure as previously reported in the lower resolution simulation.
It has to be mentioned that the computational size is $L=16 \ h^{-1}$
Mpc and that a lower box size would show a more resolved structure.
We conclude that the introduction of non-equipartition processes leads
to an out of equilibrium IGM, and that their influence is likely to be
dominant at epochs before complete reionization of the universe. This
plasma is involved in the galaxy formation process and we now turn to
the cosmological implications.

\subsection{On galaxy formation}
\label{galaxy}

Galaxy formation is closely related to the ability of gas to cool. One
necessary condition for the collapse of gas clouds is that the cooling
timescale must be less than the dynamical timescale. Cooling rates
depend on the electron density and we have seen that, in out of
equilibrium regions, the electron density can be lower by up to two
orders of magnitude than in the simulation with equipartition
processes forced. We compare the cooling timescales computed for an
out of equilibrium plasma (subscript $\mathrm{NE}$) and for a plasma
in equilibrium (subscript $\mathrm E$). In the first case the plasma
temperature is the bulk temperature, mainly the heavy particle one:
\begin{eqnarray}
\label{tcoolNE}
(t_\mathrm{cool})_\mathrm{NE} &=& \frac{3/2kn_\mathrm{tot}T_\mathrm
g}{Q_\mathrm{cool}(T_\mathrm e,x)}, \\
\label{tcoolE}
(t_\mathrm{cool})_{\mathrm E} &=&
\frac{3/2kn_\mathrm{tot}T}{Q_\mathrm{cool}(T_\mathrm e,x)}.
\end{eqnarray}
Note that in the second expression, the electron temperature in the
cooling rate term is also the temperature of the plasma, $T=T_\mathrm
e$. These timescales are normalized to the dynamical timescale:
\begin{equation}
\label{tdyn}
t_\mathrm{dyn} = \sqrt\frac{3 \pi}{32 G \rho},
\end{equation}
and are plotted in Fig.~\ref{compt_tps3} as a function of the
ionization degree and the electron temperature, since this temperature
controls the cooling rates. In these expressions $Q_\mathrm{cool}$ is
computed from equation (\ref{qcool}), $n_\mathrm{tot}$ is fixed at
$5.10^{-5} \ \textrm{cm}^{-3}$, $n_\mathrm H$ is given by
$n_\mathrm{tot}/(1+x)$ and $\rho$ is $n_\mathrm H m_\mathrm
p+n_\mathrm e m_\mathrm e$. In equation (\ref{tdyn}) $\rho$ is
multiplied by a factor of 10 to include the dark matter component. \\

Two points matching the thermodynamical state of the plasma in the two
simulations are worth mentioning. A point located at $x=5.10^{-4}$ and
$T_\mathrm e=1.5 \times 10^4$ K mimics the warm weakly ionized
baryonic matter in the outer regions of not too dense structures (such
as in Fig.  \ref{carteTS3T}) seen in the $S_{3\mathrm T}$
simulation. A point located at $x=10^{-1}$ and $T_\mathrm e=2.10^4$ K
mimics the cold partially ionized baryonic matter in similar
structures in the $S_{1\mathrm T}$ simulation.
Figure~\ref{compt_tps3} shows that the first point is in the region
$t_\mathrm{cool}>t_\mathrm{dyn}$ whereas the second point is in the
region $t_\mathrm{cool}<t_\mathrm{dyn}$.  This suggests that the out
of equilibrium plasma in the $S_{3\mathrm T}$ simulation is already
cooled in the $S_{1\mathrm T}$ simulation. Thus the cooling timescale
is expected to be longer in simulations taking into-account
non-equipartition processes. \\

We therefore estimate the cold baryonic mass fraction in the two
simulations. Figure~\ref{fmBM_BM} presents the baryonic mass fractions
computed at two redshifts as a function of the baryonic density
contrast in three ranges of temperature: $8.10^3 \leq T < 1.7 \times
10^4$ K, $1.7 \times 10^4 \leq T < 3.10^4$ K and $3.10^4 \leq T <
5.10^5$ K. First of all we notice that the plots for the $S_{3\mathrm
T}$ simulation show a warm phase ($3.10^4 \leq T < 5.10^5$ K) which is
not found in plots for the $S_{1\mathrm T}$ simulation. As explained
above, this phase in the $S_{3\mathrm T}$ simulation is cold in the
latter simulation and this is the second point to notice: the mass
fraction of cold gas ($8.10^3 \leq T < 1.7 \times 10^4$ K) is lower in
the $S_{3\mathrm T}$ simulation than in the $S_{1\mathrm T}$
simulation.  We have previously shown the evolution in redshift of the
warm and cold fractions in Fig.  \ref{glob_lcdm}. Since this cold
phase is a reservoir for galaxy formation, this analysis of the
thermodynamic properties and distributions of baryonic matter,
strongly suggests that modifications in galaxy formation are to be
expected in models with non-equipartition processes. This issue will
be quantified in a forthcoming paper focusing on simulations including
non-equipartition processes and galaxy formation (Alimi $\&$ Courty,
in preparation).

\section{Conclusions}

We have considered in this paper the influence of non-equipartition
processes on the thermodynamic properties of baryons in the universe,
using numerical simulations of large scale structure formation. In the
$S_{3\mathrm T}$ simulation each species of the plasma, electrons,
ions and neutral particles, has its own kinetic energy. The results
have been compared with the $S_{1\mathrm T}$ simulation in which
equipartition processes are forced. In the $S_{3\mathrm T}$
simulation, plasma is out of equilibrium in the outer regions of
structures where the accreted baryonic matter is shocked or
gravitationally compressed. The causes of the departure from
equilibrium depend on the interaction mechanisms and thus on the
thermodynamics of the accreted plasma. In the intra cluster medium of
massive structures, baryons are shock heated to very high temperatures
($\sim 10^8$ K) and the equipartition timescale between ions and
electrons is of the order of the Hubble time. But in low mass
structures we have pointed out that departure from equilibrium in
weakly ionized accreted plasma is driven by interactions between
electrons and neutral particles.  In that case the temperature
differences are in the range $10^4$--$10^6$ K and in the $S_{3\mathrm
T}$ simulation the plasma is found to be warmer than in the
$S_{1\mathrm T}$ simulation. We have checked that the thermodynamic
differences are not due to artificial heating. Since the energy
exchange timescale between neutral particles and electrons is
relatively short compared to the Hubble time, we have estimated
semi-analytically the relaxation timescale in weakly ionized low
temperature out of equilibrium regions. We have shown that departure
from equilibrium results from the competition between the cooling
rates of the electrons and their heating rates by heavy particles. The
relaxation timescale has been found to be long enough to yield
departure from equilibrium. Moreover a semi-analytical relation
between heavy particle and electron temperatures parameterized by the
ionization degree has been derived and our numerical results are in a
good agreement with this relation.  Finally, we have discussed the
cosmological implications of the warmer intergalactic medium in
simulations taking non-equipartition processes into account: we have
concluded that galaxy formation is expected to be modified and this
issue will be studied in detail in forthcoming papers, in terms of
galaxy properties and galaxy clustering.

\begin{acknowledgements}
  
  Numerical simulations of this paper were performed on NEC-SX5 at the
  I.D.R.I.S. computing center (France). We thank Arturo Serna and
  Gunnlaugur Bj\"ornsson for a careful reading of the paper and the
  referee, Andrea Ferrara, for his constructive comments. JMA thanks
  Laurent Moog (young student, now deceased) for fruitful discussions
  about the cosmological constant. SC acknowledges partial support
  from a special grant from the Icelandic Research Council during the
  final stages of this work.

\end{acknowledgements}

\bibliography{paper}

\end{document}